\documentclass[fleqn,usenatbib,useAMS]{mnras}

\usepackage{graphicx}	% Including figure files
\usepackage{amsmath}	% Advanced maths commands
\usepackage{amssymb}	% Extra maths symbols
\usepackage{multicol}        % Multi-column entries in tables
\usepackage{bm}		    % Bold maths symbols, including upright Greek
\usepackage{pdflscape}	% Landscape pages
\usepackage{xcolor}	    % More colors

\usepackage[T1]{fontenc}
\usepackage{ae,aecompl}

\usepackage{newtxtext,newtxmath}

\title[Relentless Transits from a Debris Disk]{Relentless and Complex Transits from a Planetesimal Debris Disk}

\author[J. Farihi et al.]{J. Farihi$^1$\thanks{E-mail: j.farihi@ucl.ac.uk},
J. J. Hermes$^2$,
T. R. Marsh$^3$,
A. J. Mustill$^4$,
M. C. Wyatt$^5$,
J. A. Guidry$^2$,
\newauthor
T. G. Wilson$^{1,6}$,
S. Redfield$^7$,
P. Izquierdo$^8$,
O. Toloza$^{3,9}$,
B. T. G\"ansicke$^3$,
\newauthor
A. Aungwerojwit$^{10}$,
C. Kaewmanee$^{11}$, 
V. S. Dhillon$^{12,13}$,
A. Swan$^1$
\medskip
\\
$^1$Department of Physics and Astronomy, University College London, London WC1E 6BT, UK\\
$^2$Department of Astronomy \& Institute for Astrophysical Research, Boston University, Boston 02215, USA\\
$^3$Department of Physics, University of Warwick, Coventry CV4 7AL, UK\\
$^4$Lund Observatory, Department of Astronomy \& Theoretical Physics, Lund University, SE-221 00 Lund, Sweden\\
$^5$Institute of Astronomy, University of Cambridge, Cambridge CB3 0HA, UK\\
$^6$School of Physics and Astronomy, University of St. Andrews, St. Andrews KY16 9SS, UK\\
$^7$Astronomy Department and Van Vleck Observatory, Wesleyan University, Middletown 06459, USA\\
$^8$Departamento de Astrof\'isica, Universidad de La Laguna, E-38206 La Laguna, Spain\\
$^9$Departamento de F\'isica, Universidad T\'ecnica Federico Santa Mar\'ia, Avenida, Valpara\'iso, Chile\\
$^{10}$Department of Physics, Naresuan University, Phitsanulok 65000, Thailand\\
$^{11}$Department of Mathematics, Naresuan University, Phitsanulok 65000, Thailand\\
$^{12}$Department of Physics and Astronomy, University of Sheffield, Sheffield S3 7RH, UK\\
$^{13}$Instituto de Astrof\'isica de Canarias, E-38205 La Laguna, Spain
}

\begin{document}

%\label{firstpage}

\maketitle

\begin{abstract}
This article reports quasi-continuous transiting events towards WD\,1054--226 at $d=36.2$\,pc and $V=16.0$\,mag, based on simultaneous, high-cadence, multi-wavelength imaging photometry using ULTRACAM over 18 nights from 2019 to 2020 March.  The predominant period is 25.02\,h, and corresponds to a circular orbit with blackbody $T_{\rm eq}=323$\,K, where a planetary surface can nominally support liquid water.  The light curves reveal remarkable night-to-night similarity, with changes on longer timescales, and lack any transit-free segments of unocculted starlight.  The most pronounced dimming components occur every 23.1\,min -- exactly the 65$^{\rm th}$ harmonic of the fundamental period -- with depths of up to several per cent, and no evident color dependence.  Myriad additional harmonics are present, as well as at least two transiting features with independent periods.  High-resolution optical spectra are consistent with stable, photospheric absorption by multiple, refractory metal species, with no indication of circumstellar gas.  {\em Spitzer} observations demonstrate a lack of detectable dust emission, suggesting that the otherwise hidden circumstellar disk orbiting WD\,1054--226 may be typical of polluted white dwarfs, and only detected via favorable geometry.  Future observations are required to constrain the orbital eccentricity, but even if periastron is near the Roche limit, sublimation cannot drive mass loss in refractory parent bodies, and collisional disintegration is necessary for dust production.
\end{abstract}


\section{Introduction}

Evidence of planetary systems orbiting hundreds of white dwarf stars is indirect but compelling, where the telltale signature of metal pollution is consistent with the accretion of rocky planetesimals for isolated stars, and indicative of closely orbiting debris compatible with planetary tidal disruption events \citep{koester2014,farihi2016a}.  In a growing number of these systems, the geochemical hallmarks of atmospheric pollution \citep{jura2014,doyle2019} are joined by irregular transit or dimming events \citep{guidry2021,vanderbosch2020,vanderbosch2021}.  In two cases, including one lacking atmospheric pollution, there may be actual giant planets in evidence \citep{vanderburg2020,gansicke2020}.

The well-studied system and prototype transiting system, WD\,1145+017, exhibits dimming events with associated periodicities concentrated near 4.5\,h, consistent with disintegrating, minor planetary bodies orbiting near the Roche limit \citep{vanderburg2015,rappaport2016}.  In addition to the variable stellar occultations, the debris disk surrounding this star is detected via infrared emission and also by evolving gaseous absorption along the line of sight \citep{xu2016,redfield2017}.  The disk has been shown to precess \citep{cauley2018} and is therefore at least mildly eccentric \citep{miranda2018,fortin2020}, and in this respect similar to other debris disks orbiting white dwarfs \citep{manser2016a,dennihy2018}.  

Eccentric orbits are in line with theoretical predictions \citep{malamud2020,nixon2020}.  It is expected that extant planetary bodies orbit outside the radius of the prior first-ascent and asymptotic giant branch stellar radii, and are perturbed into close stellar approaches by major planets via dynamical instabilities \citep{bonsor2011,debes2012a,frewen2014,smallwood2018,mustill2018}.  The process of orbital compaction and circularization is almost certainly driven by collisions \citep{malamud2021,li2021}, as Poynting-Robertson drag is far too feeble across the bulk of parameter space for white dwarf luminosities, particle sizes, and initial orbits \citep{veras2015}.  In contrast, infrared and complementary wavelength disk observations indicate that most, if not all known disks have compact radii that are in the neighborhood of the Roche limit, with no cooling age dependence on the shape of the dust spectral energy distribution \citep{farihi2009a,rocchetto2015}.

Collisions appear responsible for the infrared brightening and dimming variability that is observed nearly ubiquitously towards polluted and dusty white dwarfs \citep{swan2019a}.  Because the magnitude of infrared variability is strongest among sources where both circumstellar dust and gas are found, and yet is clearly part of a continuum of infrared changes among the larger population of dusty white dwarfs \citep{xu2014b,swan2020}, sublimation cannot be responsible for the liberation of the detected gas.  The infrared observations require both dust replenishment and removal, as both increases and decreases in dust emission have been observed \citep{farihi2018c}.  Collisions provide a mechanism for both of these processes, as well as producing gas in debris disk models \citep{kenyon2017b,malamud2021,swan2021}.  The rate of dust and gas production has direct bearing on the size distribution of particles and the largest parent bodies within evolving debris disks \citep{wyatt2008}, and ultimately for the accretion rate onto the stellar surface \citep{metzger2012,wyatt2014}.

%%% TABLE STELLAR PARAMS %%%
\begin{table}
\begin{center}
\caption{Published and calculated parameters of WD\,1054--226\label{tbl:sparams}.}
\label{tab:Tab1}
\begin{tabular}{lr}
		
\hline

Parameter					&Value\\

\hline

Spectral Type				&DAZ\\
$V$ (mag)					&16.0\\
Distance (pc)				&$36.17\pm0.07$\\
$v_{\rm tan}$ (km\,s$^{-1}$)&$52.9\pm0.1$\\				
$T_{\rm eff}$ (K)			&$7910\pm120$\\
$\log\,g$ (cm\,s$^{-2}$)	&$8.05\pm0.08$\\
$M_*$ (M$_{\odot}$)			&$0.62\pm0.05$\\
$R_*$ (R$_{\odot}$)			&$0.012\pm0.001$\\
$\log(L_*/$L$_{\odot})$		&$-3.27\pm0.09$\\
Cooling Age (Gyr)			&$1.3\pm0.2$\\

\hline

\end{tabular}
\end{center}
{\em Note}.  Stellar parameters are those derived from spectroscopy \citep{gianninas2011}, apparent brightness and spectral type from an earlier census of nearby white dwarfs \citep{subasavage2007}, and distance from {\em Gaia} EDR3 \citep{gaia2021}.  The remaining quantities are calculated, with cooling age based on white dwarf evolutionary models \citep{fontaine2001}.  
\end{table}

%%% TABLE OBSERVING RUNS %%%
\begin{table}
\begin{center}
\caption{Summary of ULTRACAM observing runs\label{tbl:obsruns}.}
\begin{tabular}{@{}ccccc@{}}

\hline

Run	&Date			&Coverage	&Phase		&Filter\\
(\#)	&        			&(h)			&Range		&Set\\
				
\hline

1	&2019 Mar 24		&8.4			&0.00--0.34	&$ugr$\\
1	&2019 Mar 25		&8.3			&0.96--0.29	&$ugr$\\
1	&2019 Mar 26		&8.3			&0.91--0.25	&$ugr$\\
1	&2019 Mar 27		&7.6			&0.89--0.19	&$ugr$\\
1	&2019 Mar 28		&4.9			&0.98--0.17	&$ugr$\\
1	&2019 Mar 29		&3.7			&0.80--0.95	&$ugr$\\
1	&2019 Mar 30		&4.3			&0.91--0.08	&$ugr$\\
1	&2019 Mar 31		&8.0			&0.72--0.03	&$ugr$\\
2	&2019 Apr 10		&2.4			&0.50--0.59	&$ugr$\\
3	&2019 May 03		&6.0			&0.36--0.60	&$ugr$\\
3	&2019 May 04		&6.4			&0.30--0.56	&$ugr$\\
3	&2019 May 05		&6.2			&0.25--0.50	&$ugr$\\
3	&2019 May 07		&2.6			&0.17--0.27	&$ugr$\\
4	&2020 Feb 28		&9.3			&0.12--0.49	&$ugr$\\
4	&2020 Feb 29		&9.6			&0.07--0.46	&$ugr$\\
4	&2020 Mar 01		&9.4			&0.04--0.42	&$ugi$\\
4	&2020 Mar 02		&9.5			&0.99--0.37	&$ugi$\\
4	&2020 Mar 03		&9.3			&0.95--0.32	&$ugi$\\

\hline

\end{tabular}
\end{center}
{\em Note}.  The phases shown are for the 25.02\,h, fundamental period. 
\end{table}

It is not yet known whether white dwarf transit detections are related to dynamical activity such as ongoing collisions or catastrophic fragmentation, or purely the result of viewing geometry.  A heightened dynamical state of dust and gas production may lead to a dramatic increase in disk scale height (e.g.\ \citealt{kenyon2017a}), and favor detection via increased transit cross-section.  Alternatively, the dimming may simply be a result of favorable inclination with respect to the observer, and where the inferred scale height of transiting structures are typical of all circumstellar disks orbiting white dwarfs.  X-ray upper limits for WD\,1145+017 do not favor a heightened state of dust and gas production in that system \citep{farihi2018a,rappaport2018a}.

The number of known white dwarfs with transits has grown; the second was reported in 2020 \citep{vanderbosch2020}, and seven total are now known \citep{guidry2021}.  There appears to be a variety of orbital periods and transit depths, where only WD\,1145+017 exhibits periodicity near that expected of the Roche radius.  ZTF\,J013906.17+524536.9 (\citealt{vanderbosch2020}; hereafter ZTF\,0139) has deep dimming events similar to the prototype, but the orbital period of the material is orders of magnitude larger where $p\sim100$\,d.  This transiting material may partially orbit within the Roche limit, as transits are more probable near periastron, and tidal disruption is likely to be ultimately responsible for the evolving clouds of orbiting debris.  Nevertheless, the source of the planetary material is clearly further out, and on a scale of au rather than solar radii.  ZTF\,J032833.52-121945.3 (\citealt{vanderbosch2021}; hereafter ZTF\,0328) presents two independent but notably similar, recurrent transit signals, where both are $p\sim10$\,h, and thus closer to the prototype, but the transit depths in this system are on the order of 10\,per cent.

These transiting systems provide further signposts on the road from planetesimal to debris disk to atmospheric pollution; e.g.\ the size of the occulting bodies, constraints on compositions and particle sizes, orbital periods and source regions.  Using high-cadence imaging photometry, this paper reports the discovery and detailed characterization of transiting events towards the polluted white dwarf WD\,1054--226 (=LP\,849-31), a 16$^{\rm th}$ magnitude star within the local 40\,pc volume.  The observations and data are presented in Section~2, the time-series, spectroscopic, and infrared data analysis are presented in Section~3, and inferences based on the data are discussed in Section~4.  An outlook is provided in Section~5.

\section{Observations and data}

WD\,1054--226 is a relatively cool and nearby, hydrogen atmosphere, white dwarf star that displays a classical DA spectrum \citep{subasavage2007}, and its properties are summarized in Table~\ref{tbl:sparams}.  The {\em Gaia} EDR3 parallax places the source at $d=36.17\pm0.07$\,pc, and with $G=16.0$\,mag \citep{gaia2021} it is brighter than other transiting white dwarfs by a factor of around two or more.  At modest resolution, the stellar spectrum exhibits a typical signature of external pollution by heavy elements via Ca\,{\sc ii} K absorption, yielding a spectral type of DAZ with $T_{\rm eff}=7910$\,K as determined by atmospheric modeling \citep{gianninas2011}.  At higher spectral resolution, there are photospheric absorption lines from multiple metal species of Mg, Al, Ca, and Fe, at fairly typical abundances relative to the known population of similarly polluted stars \citep{vennes2013}.  WD\,1054--226 was targeted as part of an ongoing effort to identify transit events towards metal-enriched white dwarfs.

\subsection{Time-Series Photometry with ULTRACAM}

The primary photometry and light curves for WD\,1054--226, including the discovery of the transiting features, were obtained with ULTRACAM, a triple-channel, frame-transfer CCD imaging camera with 24\,ms dead time between exposures \citep{dhillon2007}.  The observations were carried out at La Silla Observatory on the NTT telescope, using filters similar to standard $ugri$ bandpasses but with higher throughput.  The design of ULTRACAM allows the use of three filters simultaneously, which were $ugr$ for all nights excepting 2020 March 1--3 where $ugi$ filters were in place.  High-speed imaging was performed with exposure times per frame of 6.03\,s in each band, with the bluest channel typically co-added every three frames to overcome readout noise.  Observations were made during 2019 March, April, May, and 2020 February-March under variable conditions with coverage and filters as documented in Table~\ref{tbl:obsruns}.  Useful light curves were obtained on 18 nights, where the total time on the science target is approximately 124.2\,h.

All science frames were corrected for bias, and flat fielded using normalized sky flats obtained in a continuous spiral (to remove stars) during evening twilight each night.  All three arrays had readout noise that was typically $2-3$\,ADU in the bias frames taken each night.  Differential photometry of the science target was measured relative to nearby field stars, using instrument-dedicated software to measure fluxes using apertures. The target and comparison star apertures, which tracked the position of the stars, had radii that were scaled to $2\times$ the mean full width at half maximum of the stellar profiles for each exposure.  The sky annuli were fixed to span the region 8.75--15.75\,arcsec from the stars, where a clipped mean was used to determine the background.  Signal-to-noise (S/N) values for the science target in an individual exposure were typically near or above 200 for the red ($r$ or $i$ band) and green ($g$ band) channels, and around 150 for the blue ($u$ band) channel.  The entire ULTRACAM dataset has been reduced using a uniform set of comparison stars, where in the end, two field stars were used for normalization; these are Gaia\,EDR3\,3549471169391629568 with $G=13.6$\,mag, and Gaia\,EDR3\,3549473402774675712 with $G=14.5$\,mag (cf.\ $G=16.0$\,mag for WD\,1054--226).  

Supplemental observations were obtained with ULTRASPEC \citep{dhillon2014} mounted on the 2.4m Thai National Telescope on the nights starting 2019 March 25 and 26.  The runs were of duration 5.2 and 4.7\,h, and exposure times of 8.1 and 18.4\,s were used on the two nights, respectively.  Thailand is $168\degr$ of longitude east of Chile, so these observations took place largely during Chilean daytime, with the first run obtained in between the first two nights of ULTRACAM data.  A broad filter spanning $u$, $g$, and $r$ was used for maximum signal, although the data were badly affected by clouds.  The data were reduced in the same manner as for ULTRACAM, and despite poor weather, the ULTRASPEC data proved helpful in eliminating possible period aliases (as described later).

All timestamps were converted to Barycentric Julian Day (BJD) using Barycentric Dynamical Time (TDB), following published methods that yield a precision better than 1\,ms \citep{eastman2010}.

\subsection{{\em TESS} Data}

The {\em Transiting Exoplanet Survey Satellite (TESS)} collected 120\,s cadence observations of WD\,1054--226 in Sector 9 (GI program G011113) as target TIC\,415714190 ($T$=15.8\,mag).  These observations span 2019 February 28 to March 25, and cover 24.2\,d with a duty cycle of 91.4\,per cent.  Almost all of the {\em TESS} coverage is a continuous series of 120\,s exposures, with a two-day gap in the middle caused by the mid-sector data downlink.  Additionally, {\em TESS} observed WD\,1054--226 just over two years later in Sector 36 with the fastest cadence possible, using 20\,s exposures (GI programs G03124 and G03207).  These observations span 2021 March 7 to April 1, and cover 23.7\,d with a duty cycle of 89.0\,per cent.

Light curves were extracted by the Science Processing Operations Center (SPOC) pipeline, which corrects for long-term instrumental systematics \citep{jenkins2010,jenkins2016}.  The SPOC-processed light curve has not been significantly altered, except by application of a $5\upsigma$ clip to remove the most discrepant points.  Crowding in the large {\em TESS} pixels is problematic for such faint stars, and in both sectors the SPOC pipeline estimates that the target contributes only half the total flux in the extracted aperture. This flux dilution has automatically been compensated for in the {\tt PDCSAP\_FLUX} values from the SPOC pipeline.

The average point-to-point scatter for the 120\,s observations in Sector 9 is roughly 9.8\,per cent, and for the 20\,s observations in Sector 36 it exceeds 12.4\,per cent. This large scatter prevents the direct identification of any features in the light curve, but the dense, long-baseline observations are highly useful for periodogram analysis.

%%% FIGURE UCAM 18 NIGHTS PHASED %%%
\begin{figure*}
\includegraphics[width=1.0\linewidth]{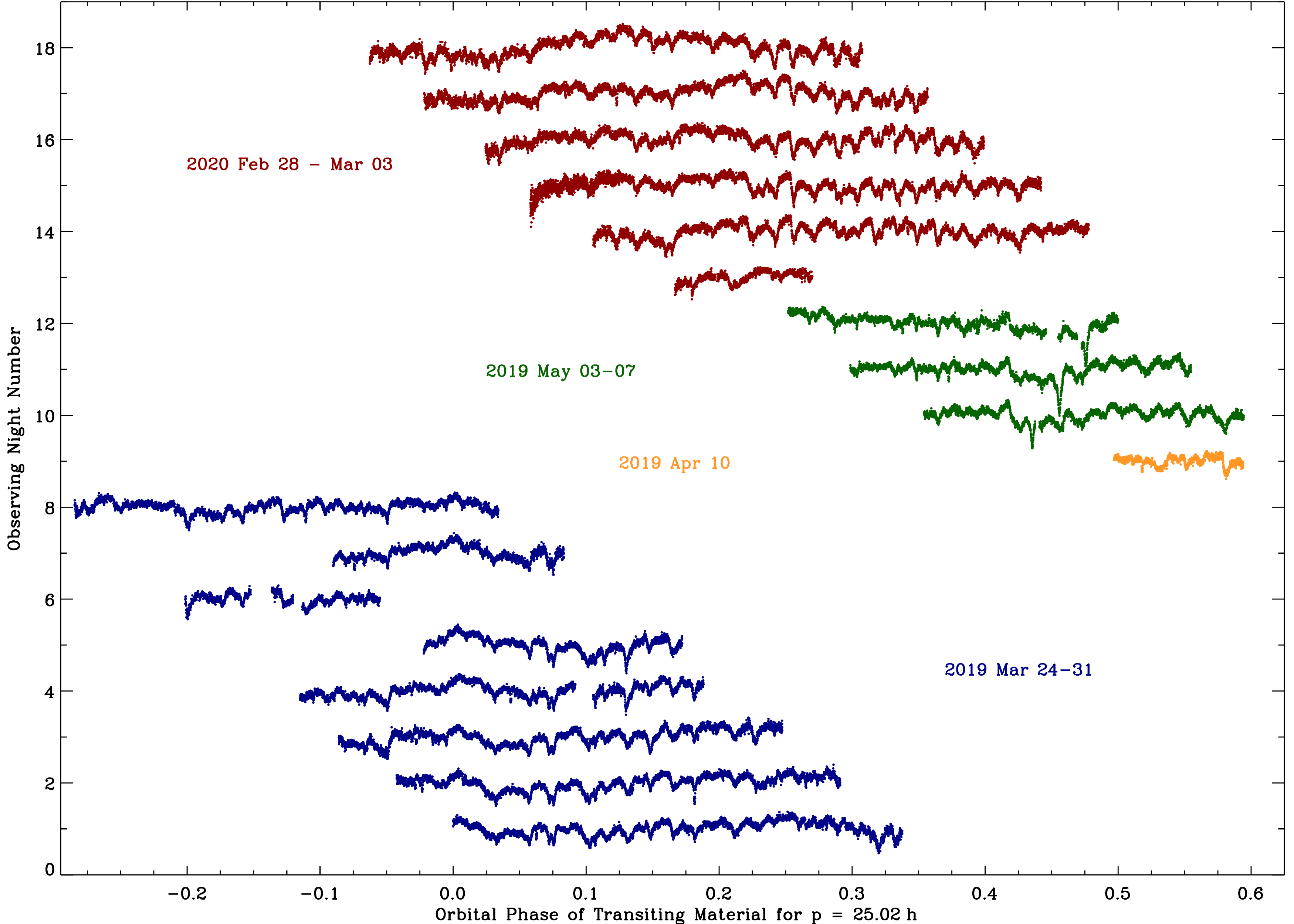}
\caption{All light curves obtained with ULTRACAM, shown here as the sum of the $g$- and $r$-band data, phased to the 25.02\,h period.  The light curves were first normalized to one, then multiplied (stretched) by a factor of ten for visibility, then offset by integer values so that the $y$ axis corresponds to the order in which the light curves were obtained by night (note there are gaps between, and also within, observing runs).  The four distinct campaigns are delineated by separate colors and labelled.  Within each observing run lasting a few to several nights, the recurrence of virtually all transit features can be discerned.  A closer view of the first three nights is provided in Figure~\ref{fig:nights3}, and all ULTRACAM light curves are shown in the Appendix, including examples in each bandpass (Figure \ref{fig:firstlastcol}).
\label{fig:phase18}}
\end{figure*}

\subsection{High-Resolution Spectroscopy with UVES}

WD\,1054--226 was observed with the Ultraviolet and Visual Echelle Spectrograph (UVES; \citealt{dekker2000}) using four observing blocks carried out on three nights, 2019 May 10, 28, and 30 based on an ESO DDT award.  Two standard dichroic settings were used to capture the bulk of optical wavelengths, with one setting at $\uplambda_{\rm c} = 3460/5640$\,\AA, and the other centered on $3900/7600$\,\AA.  All spectra were acquired using the 1.0\,arcsec slit and $2\times2$ binning on the arrays, for a nominal resolving power $R\approx40\,000-50\,000$.  Observations with each of these instrument settings were performed twice, with 1.0\,h individual exposures and a total of 4.0\,h per setup.  In this way, each wavelength is covered in either four or eight total exposures.

The raw science and calibration frames were reduced using the standard {\sc reflex} UVES pipeline.  Following best practices and parameter optimizations recommended in the {\sc reflex} \citep{freudling2013} documentation, the spectra were flat fielded, bias and dark subtracted, and wavelength calibrated.  Optimal extraction was conducted utilizing the various profiles available in the pipeline to produce the highest S/N spectra.  The observations were taken in pairs with identical observational setups, were combined via weighted average, and subsequently the spectra were normalized in each arm.  Two pairs of (four total) observations were obtained for both instrumental setups, yielding two co-added spectra for each setup at the four distinct epochs.  For the $3460/5640$\,\AA \ setup, the calculated S/N for the two co-added spectra are 28 and 30 in the blue, and 63 and 67 in the red, as measured in the ranges $3500-3550$\,\AA \ and $5450-5550$\,\AA, respectively.  For the $3900/7600$\,\AA \ setup, the S/N in the $4420-4470$\,\AA \ and $6000-6100$\,\AA \ regions are 44 and 50 in the blue, and 52 and 62 in the red.

\subsection{Infrared Photometry with warm {\em Spitzer} IRAC}

WD\,1054--226 was observed with the {\em Spitzer Space Telescope} \citep{werner2004} on 2019 May 11 in both the 3.6\,$\upmu$m and 4.5\,$\upmu$m channels of warm IRAC \citep{fazio2004}.  The target was observed using 30\,s exposures in the medium step size, cycling dither pattern, with 42 frames acquired and just over 20\,min on-source at each infrared wavelength.  Both 0.6\,arcsec\,pixel$^{-1}$ mosaics and 1.2\,arcsec\,pixel$^{-1}$ individual basic calibrated data frames were processed with pipeline S19.2.0 and downloaded from the archive.  The source is detected at S/N $\approx150$ in channel 1, and S/N $\approx100$ in channel 2.

%%% FIGURE UCAM FIRST + LAST 3 NIGHTS %%%
\begin{figure*}
\includegraphics[width=1.0\linewidth]{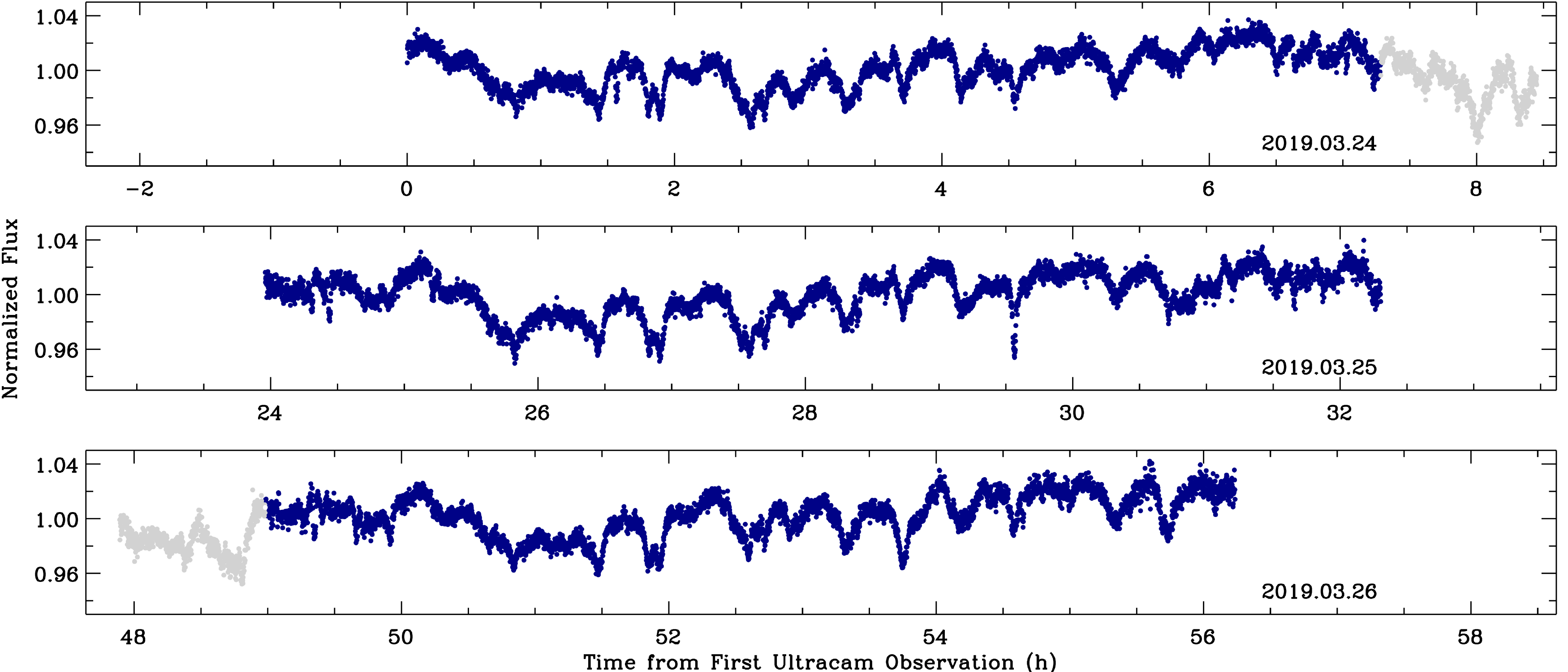}
\caption{The first three nights of ULTRACAM observations for WD\,1054--226, shown as an average of the $g$- and $r$-band fluxes.  The $x$ axes of the plots are structured so that the same orbital phase occurs at the same $x$ position, and highlights the vertical alignment of the recurring features every 25.02\,h.  The grey points are portions of the light curves with no corresponding phase coverage in the other two nights.  The nearly exact replication of the transit structures each night is striking.
\label{fig:nights3}}
\end{figure*}

The target flux was measured both using aperture photometry in {\sc iraf} on the image mosaics, and with the {\sc apex} module within {\sc mopex} to perform point-spread function fitting photometry on individual frames, all following observatory recommendations.  There are two neighboring and fainter IRAC sources located between 5.1 and 5.4\,arcsec of the white dwarf.  Aperture photometry with $r=3$ native pixels (3.6\,arcsec) yielded fluxes that are 2--4\,per cent higher than with $r=2$ native pixels (2.4\,arcsec), suggesting negligible flux contamination from the nearby sources in the smaller apertures.  The resulting mean fluxes for the 42 measurements on the individual frames agree with the smaller aperture, mosaic photometry to within 0.5\,per cent, and thus the latter values were adopted.  Because both measurements have S/N $\ga100$, the uncertainty is dominated by an assumed 5\,per cent calibration error.  The IRAC fluxes are $212\pm11\,\upmu$Jy and $134\pm7\,\upmu$Jy at 3.6\,$\upmu$m and 4.5\,$\upmu$m, respectively.

\section{Data Analysis}

The following sections outline the analyses performed on each of the datasets, with some basic inferences.

%%% FIGURE UCAM COADDS %%%
\begin{figure*}
\includegraphics[width=1.0\linewidth]{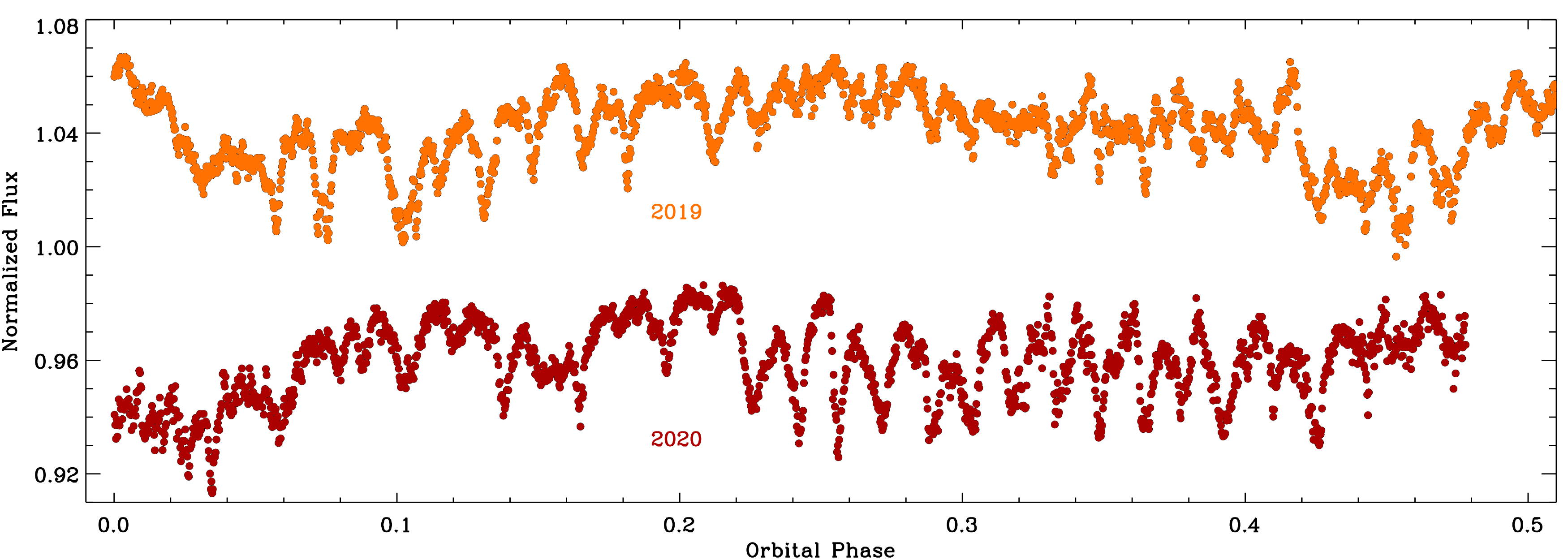}
\caption{Combined and phase-folded light curves for all 13 nights of ULTRACAM data in 2019, and all 5 nights of data in 2020.  The fluxes are phased with respect to the 25.02\,h period, where zero phase in both cases corresponds to the first exposure obtained on 2019 March 24 ($T_0=2458567.49292$\,BJD$_{\rm TDB}$).  Both plotted curves are the result of co-adding and re-sampling nightly data into 4000 evenly spaced phase bins, which corresponds to a reduction in temporal sampling from 6\,s to 23\,s.  The individual transiting features are remarkably coherent in both shape and phase over at least several nights, but their morphology can change drastically from one year to the next.  Both curves shown are the average of either $g$- and $r$-band, or $g$- and $i$-band data.
\label{fig:coadd}}
\end{figure*}

\subsection{ULTRACAM: 25\,h repeating structures}

There are a number of striking features in the ULTRACAM light curves, where the full dataset is plotted in Figure~\ref{fig:phase18} (the nightly light curves are also displayed individually in the Appendix).  One of the notable light curve characteristics is that there are no confidently identified sections that are baseline, out-of-transit starlight; not on any night, nor in any bandpass.  While there are some light curve sections that appear nearly flat for a short time, there are always higher flux peaks found on the same night.  The dimming events are modest in depth, and while it is not possible to precisely establish a reference point for transit depth, the components typically exhibit a drop of only a few per cent relative to the flux both pre- and post-dimming.  There are a small number of features that have a depth greater than around 5\,per cent from the surrounding, quasi-baseline flux.

The nightly ULTRACAM light curves illustrate that the dimming structures repeat every 25.02\,h.  Shorter-period aliases of 12.5\,h, and 8.3\,h were initially ruled out with data taken from a different longitude using ULTRASPEC (not shown here), and the fact that some light curves span over 9\,h duration.  The 25.02\,h period seen in ULTRACAM has been subsequently confirmed by {\em TESS} (see below) to be the fundamental recurrence timescale in WD\,1054--226.

While there are some modest changes in the precise shape of some transit components, the entire light curve structure repeats with remarkable fidelity from night-to-night.  A more detailed look at this recurrence behavior is shown in Figure~\ref{fig:nights3}, which plots a close-up of the first three nights of ULTRACAM data, and where many individual transiting components can be co-identified between nights.  

A notable and easily recognized recurring feature, for example, is the double-dip structure that occurs just before hours 2, 27, and 52 in Figure~\ref{fig:nights3}.  Using this structure as a visual anchor, the adjacent, antecedent and subsequent transits can all be seen to have the same morphology over all three nights.  The same is true for any portion of the light curve on any set of a few to several consecutive nights.  In fact, the aforementioned, double-dip structure recurs on all nights of the first observing run, where that orbital phase was observable based on weather, with six manifestations in total during the eight-night run.  This feature first appears on 2019 March 24, and last appears March 30, where on the eighth consecutive night, the corresponding phase is beyond morning twilight (see Figure~\ref{fig:phase18} and Appendix).

The consecutive, individual dimming events occur with a cadence of 23.1\,min, which can be visualized in Figure~\ref{fig:nights3} or in any of the individual light curves shown in the Appendix, but is more clearly seen via the co-added light curves and periodogram analysis (see below).  Because the 25.02\,h period is essentially derived by lining up all the individual, 23.1\,min events, they form an exact ratio, and the higher-frequency individual events are thus a harmonic of the fundamental period.

In Figure~\ref{fig:coadd}, the data taken in 2019 and 2020 are re-sampled into 4000 total, evenly distributed phase bins, and separately co-added at the same orbital phase.  This re-sampling reduces the temporal cadence to around 23\,s.  The bulk of phase bins include data that span several consecutive nights, yet preserve remarkably sharp features in the combined data, and this is only possible if the 23.1\,min transiting structures are stable over this time span.  Comparing the two separate years, the structures change substantially, and based on these as well as the individual light curves, it is clear that distinct transits that repeat on consecutive nights are nevertheless morphologically evolving.  The ULTRACAM observing runs are never longer than several nights with excellent temporal coverage, so the timescale for significant changes in the light curve features are constrained to be more than several nights and less than roughly one year (i.e.\ somewhere between several orbits and a few hundred orbits).  It thus appears to be the case that the underlying transit structures, and the occulting bodies and debris responsible, persist for at least several orbits.

%%% FIGURE TESS LSPS %%%
\begin{figure*}
\includegraphics[width=1.0\linewidth]{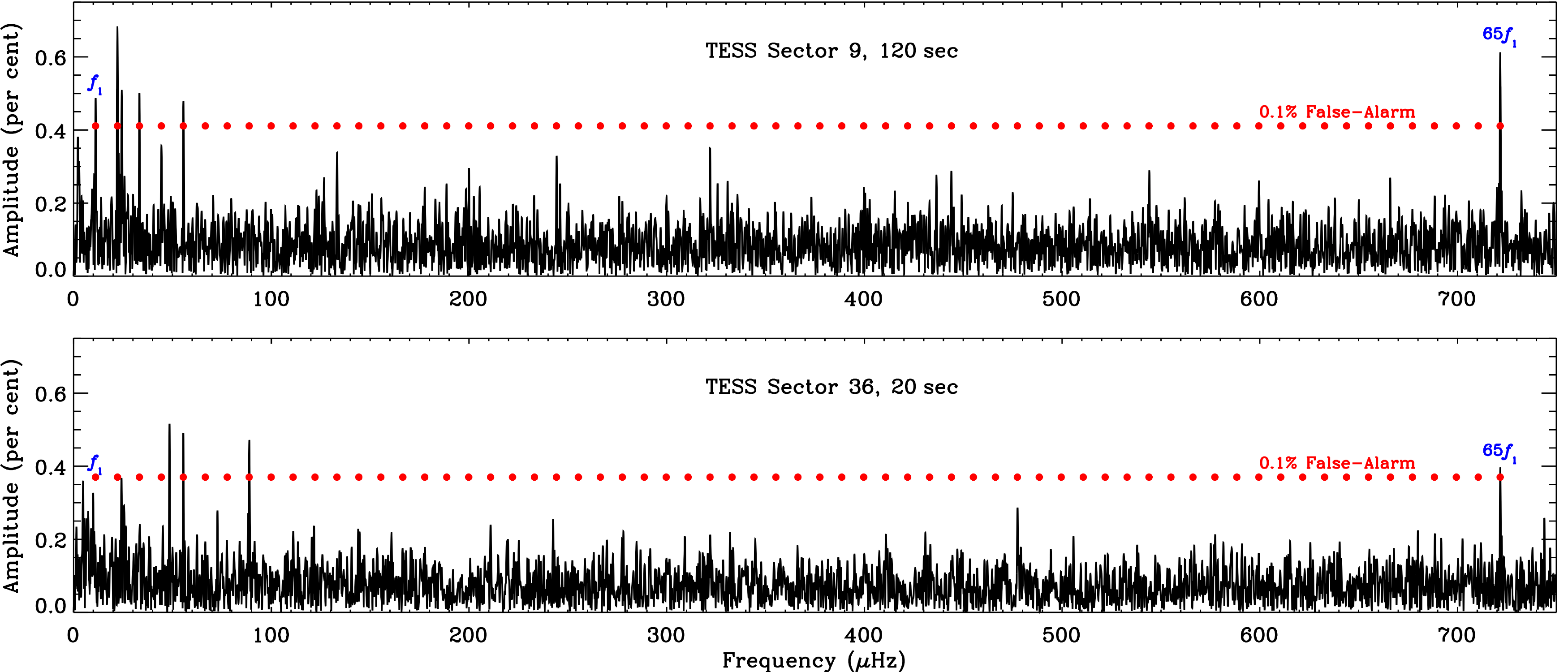}
\caption{Lomb-Scargle periodograms of the {\em TESS} data collected in Sector 9 (120\,s cadence, upper panel), and the data collected in Sector 36 (20\,s cadence, lower panel).  Many harmonics of the dominant, 25.02\,h period ($f_1 = 11.101646$\,$\upmu$Hz) are present; all integer multiples of this fundamental frequency are marked with red dots and amplitudes at the 0.1\,per cent false-alarm probability threshold, above which any peaks are formally significant.
\label{fig:lsps}}
\end{figure*}

\subsection{{\em TESS}: confirmation of the 25\,h orbit, a strong 65$^{\rm th}$ harmonic, and independent signals}

The repetition of the complex dimming event pattern, on a period of 25.02\,h as seen by ULTRACAM, inevitably implies a large series of harmonics of the fundamental frequency.  This is simply a result of Fourier series, where a multitude of sinusoids are necessary to reproduce such a spectacularly non-sinusoidal pattern.  As described below, both ULTRACAM and {\em TESS} corroborate the 25.02\,h orbital period of the transiting debris, where all light curve data reveal notable strength in the 65$^{\rm th}$ harmonic.  Additionally, {\em TESS} uncovers evidence for a second period near 11.4\,h.

The nearly continuous {\em TESS} observations provide a reliable census of the periodicities present in WD\,1054--226, free from aliases caused by gaps in the ground-based data collection.  These data can be exploited by constructing a Lomb-Scargle periodogram, which is calculated here using the software package {\sc Period04} \citep{lenz2005}.  The frequencies of the {\em TESS} data can be determined to high precision, using uncertainties derived from a linear least-squares fit \citep{montgomery1999}.  An annotated periodogram of all {\em TESS} data can be found in Figure~\ref{fig:lsps}.

Potential periods longer than 2\,d have been ignored, based on the likelihood that the longest-period signals correlate with orbital motion of the spacecraft or other systematics.  Following the method outlined in \citet{hermes2015}, a 0.1\,per cent false-alarm probability is computed by keeping the same timestamps, but randomly shuffling the fluxes to create 10\,000 synthetic light curves.  The highest-amplitude peak in the periodogram of each synthetic dataset is noted, and also the amplitude above which 99.9\,per cent of the datasets do not have a peak.  Using these, any peaks with amplitudes exceeding the 0.1\,per cent false-alarm probability are considered significant; these amplitudes are 0.411\,per cent in Sector 9, and 0.370\,per cent in Sector 36.  These significance thresholds are marked as red dotted lines in Figure~\ref{fig:lsps}.

There are many signals in the {\em TESS} periodogram of Sector 9 that are simply integer harmonics of the recurrent 25.02\,h (11.1\,$\upmu$Hz) period in the ULTRACAM data.  More importantly, in Sector 9 there is a significant peak at an apparently unrelated frequency of $24.391\pm0.038$\,$\upmu$Hz, which corresponds to $11.388\pm0.018$\,h (cf.\ the strongest signal in the periodogram of $12.544\pm0.015$\,h, the 2$^{\rm nd}$ harmonic of the fundamental).  More than two years later, {\em TESS} re-observed WD\,1054--226 in Sector 36 with higher cadence, providing a lower significance threshold in these data compared to Sector 9 (and again with many harmonics of the fundamental frequency).  The Sector 36 data also reveal a significant frequency at a value apparently unrelated to the 25.02\,h period.  This independent signal is the highest in the Sector 36 periodogram, at $48.521\pm0.033$\,$\upmu$Hz ($5.7249\pm0.0039$\,h), and is consistent with being a harmonic of another significant peak in the periodogram at $24.223\pm0.042$\,$\upmu$Hz ($11.467\pm0.020$\,h).  
This Sector 36 signal differs by only around $2\upsigma$ from the $11.388\pm0.018$\,h signal present in Sector 9, and these two independent detections suggest there is at least one additional periodicity, independent from the 25.02\,h fundamental signal, and that is physically associated with the transiting debris clouds surrounding WD\,1054--226.  

A likely alias of the 11.4\,h periodogram signal, near 22.9\,h, is found in the ULTRACAM data by visual co-identification of a repeating, individual transit feature (see Section~3.3).  If 22.9\,h were in fact the true period, then its 2$^{\rm nd}$ harmonic would be expected near 24.2\,$\upmu$Hz (11.5\,h), and a 3$^{\rm rd}$ harmonic should appear around 36.3\,$\upmu$Hz (7.6\,h); the former is present and significant, but there is no signal corresponding to the latter.  These facts, together with the strongest signal in the periodogram at 48.52\,$\upmu$Hz (5.72\,h) -- consistent with the 2$^{\rm nd}$ harmonic of 24.2\,$\upmu$Hz (11.5\,h) -- implies that the 11.4\,h signal is the true independent period, and the 22.9\,h repeating transit feature is an alias.

Both the Sector 9 and 36 data reveal the fundamental recurrence timescale (and via multiple harmonics) seen first in the ULTRACAM data, and the 65$^{\rm th}$ harmonic (that occurs at a period of 23.0966\,min) has a formally significant amplitude in all datasets (Figure~\ref{fig:lsps}).  This can be understood readily as follows: the individual dimming events have ingresses and egresses that occur every 23.1\,min, yet each temporally aligns with its own idiosyncratic counterpart every 25.02\,h.  Therefore, mathematically speaking, it is unavoidable that there will be an exact commensurability of the fundamental period with the 23.1\,min signal, and they cannot be independent.

Although the physical nature of the 65$^{\rm th}$ harmonic is as yet unknown, it can be used to refine the fundamental period by performing a linear least-squares fit to all {\em TESS} and ULTRACAM data, using the periodogram peak from Sectors 9 and 36 as an initial guess.  Such a calculation assumes a unique phasing, which is uncertain, and so instead the nearest convincing alias in the {\em TESS} and ULTRACAM periodogram of the 65$^{\rm th}$ harmonic was used for a more robust determination.  This leads to a marginally more conservative estimate of the error than assuming a unique phasing, and yields $25.0213147\pm0.0000051$\,h.

\subsection{ULTRACAM: orbiting drifters and transients}

In this section, evidence is presented for two additional orbital periods based on features in the ULTRACAM data, where these repeating transit features were seen to recur on timescales within several per cent of the 25.02\,h orbital period.  However, as noted earlier, one of these possesses a 22.9\,h periodicity that is consistent with an alias of the formally significant {\em TESS} signal at 11.4\,h seen in both Sector 9 and 36.  The other independent period does not have a counterpart in any periodogram, but it should be acknowledged that it could also be an alias of a shorter period.

On 2019 May 4, the ULTRACAM observers noted a prominent transit seen in the light curve of the previous night {\em that did not repeat at the expected time}, but instead appeared roughly half an hour later (see Figures \ref{fig:nights18} and \ref{fig:resids18}, panels dated 2019 May 3--5).  Weather was poor on the following night of 2019 May 5, but again this distinct feature -- dubbed a drifter -- is seen around 30\,min later than expected based on the 25.02\,h period.  In order to highlight this outstanding event, and to discover potential dimming events of a transient nature, or with periodicities distinct from the 25.02\,h period, the co-added and phase-folded light curves were subtracted from the nightly data to create light curve residuals, part of which are highlighted in Figure~\ref{fig:drift}.

The full set of light curve residuals are shown in the Appendix, where, overall, these data demonstrate that multiple dimming structures persist after removal of the primary recurring signals.  There are a number of dips that are either transient, or cannot be confidently shown to recur, which are thus challenging to quantify.  It is important to note that the process of creating the residuals can only effectively remove recurring signal where multiple-night coverage is available at a given phase.  Thus, some phases will suffer artefacts in this process, especially where coverage -- either in the light curve or subtracted data -- includes only a single night.  Thus, not all light curve residual structures are real.

At least two drifting transits may be seen to recur over three or more nights.  Figure~\ref{fig:drift} highlights these two repeating events, one apparently recurring more rapidly (left panel), and the other apparently recurring more slowly (right panel), than the main periodic set of features.  The faster drifting component shown in Figure~\ref{fig:drift} (and possibly a weaker dip not shown, but present in the Appendix plots) is estimated to have a period of 22.9\,h and thus co-identified with the {\em TESS} periodogram signal near 11.4\,h.  This transit event is seen in at least four nights of ULTRACAM data, and if the true period is twice as fast, then it persisted for at least eight orbits.  The more slowly drifting feature, and the strongest transit in the entire ULTRACAM dataset, has no periodogram peak in any data, but has an observed period of 25.5\,h; shorter-period aliases for this transit feature cannot yet be ruled out.

%%% FIGURE DRIFTERS %%%
\begin{figure*}
\includegraphics[width=1.0\linewidth]{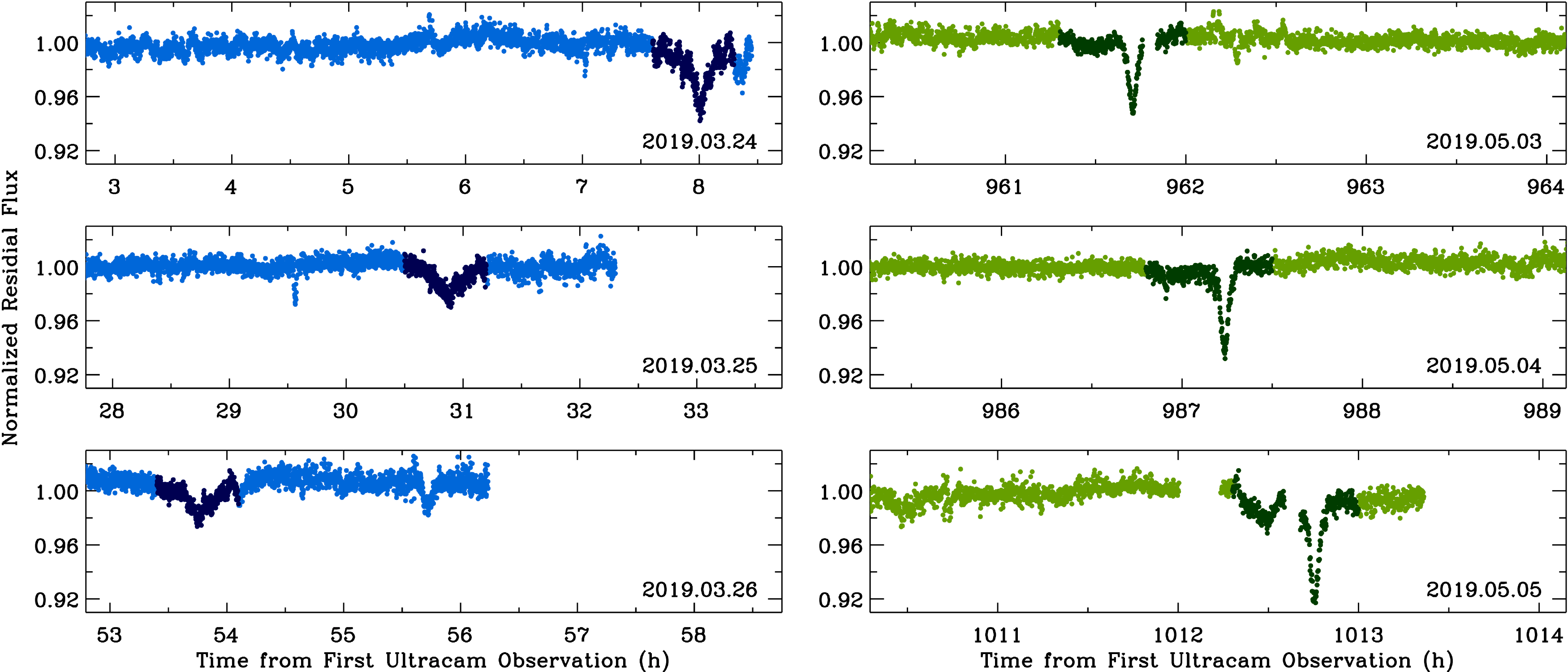}
\caption{An example of ULTRACAM, summed $g$- and $r$-band, light curve residuals that result from the removal of the phase-folded and co-added signal shown in Figure~\ref{fig:coadd}.  The plot $x$ axes are constructed analogously to Figure~\ref{fig:nights3}, where the plotted flux residuals are vertically aligned at the same orbital phase, each 25.02\,h from the previous night.  This plot shows the two most prominent residuals -- highlighted in darker color -- that recur on timescales that are distinct from the primary orbital frequency.  On the left (in blue) is a transit component that recurs more rapidly, and arriving earlier by roughly 2.1\,h, and on the right (in green) is a component that recurs more slowly, and arriving later by around 0.5\,h (see text).  Over the entire set of residual flux light curves (see Appendix), there are many transient features that cannot be confidently re-identified in adjacent nights.  It is important to note that not all light curve phases have been effectively co-added and removed in these residuals, owing to a lack of coverage at some phases (e.g.\ forward of 7\,h on 2019 March 24, where the drifter first appears).
\label{fig:drift}}
\end{figure*}

\subsection{ULTRACAM: lack of color during transits}

Visual examination of the raw, normalized flux ratios, for any pair of bandpasses, reveals that they are flat and featureless across all nights.  To examine the data for color trends in further detail, the light curves taken the first night (2019 March 24) were fitted with a five-knot, least-squares-fit, cubic spline, (with knots distributed evenly), which was then subtracted from the data, resulting in $\Delta m$ values for the light curves in all three filters.  This step guards against hourly or longer-term variations, driven by telluric extinction color terms with respect to the comparison star.  Next, a straight line was fitted to the $\Delta r$- vs. $\Delta g$-band data, where the gradient implies that the dips in WD\,1054--226 have an extinction coefficient of the form $\Delta(g-r)/\Delta g=-0.017\pm0.010$.  This can be compared to a typical interstellar extinction $\Delta(g-r)/\Delta g=0.287$ \citep{wang2019b}.

\begin{figure}
\includegraphics[width=1.0\linewidth]{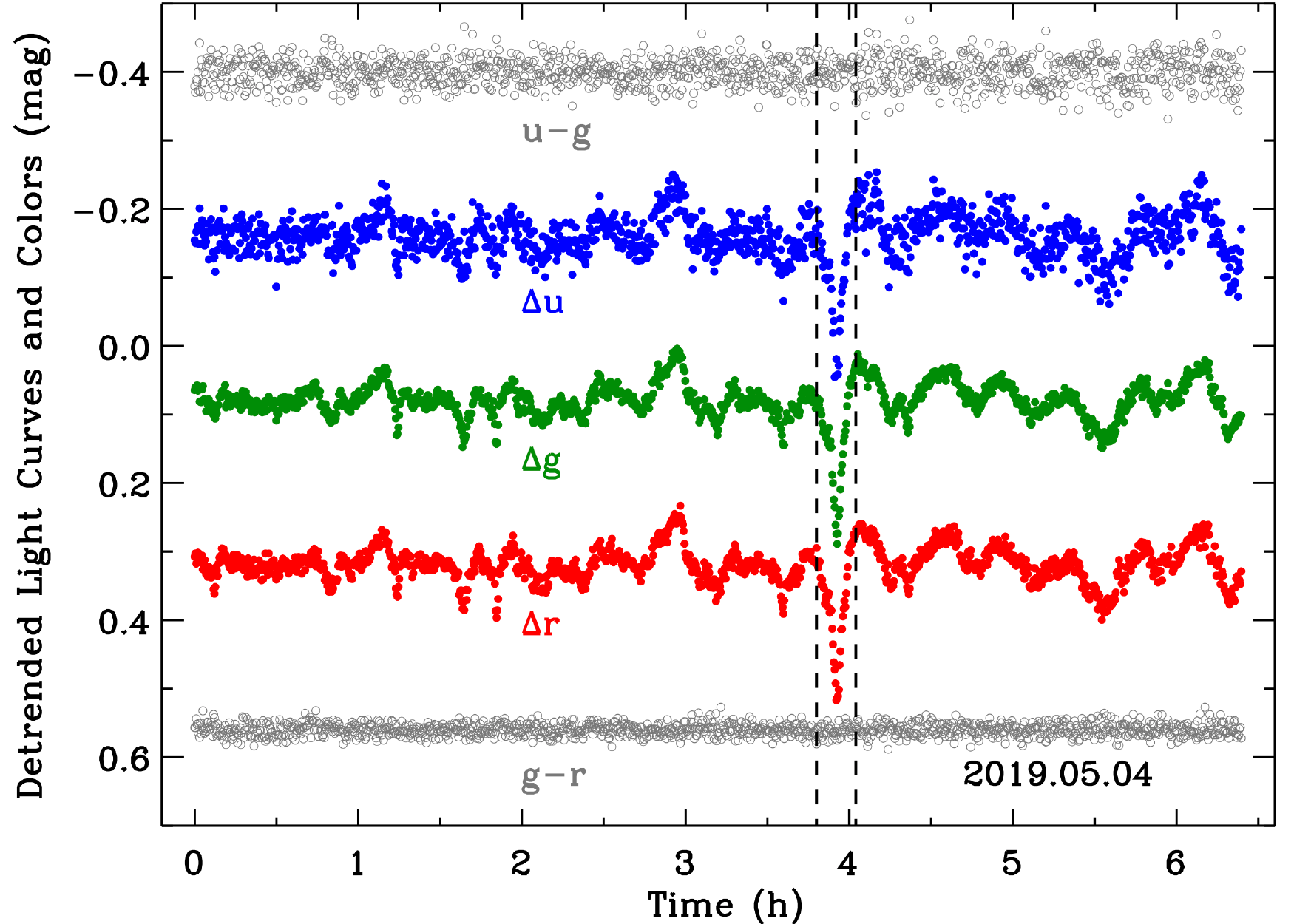}
\includegraphics[width=1.0\linewidth]{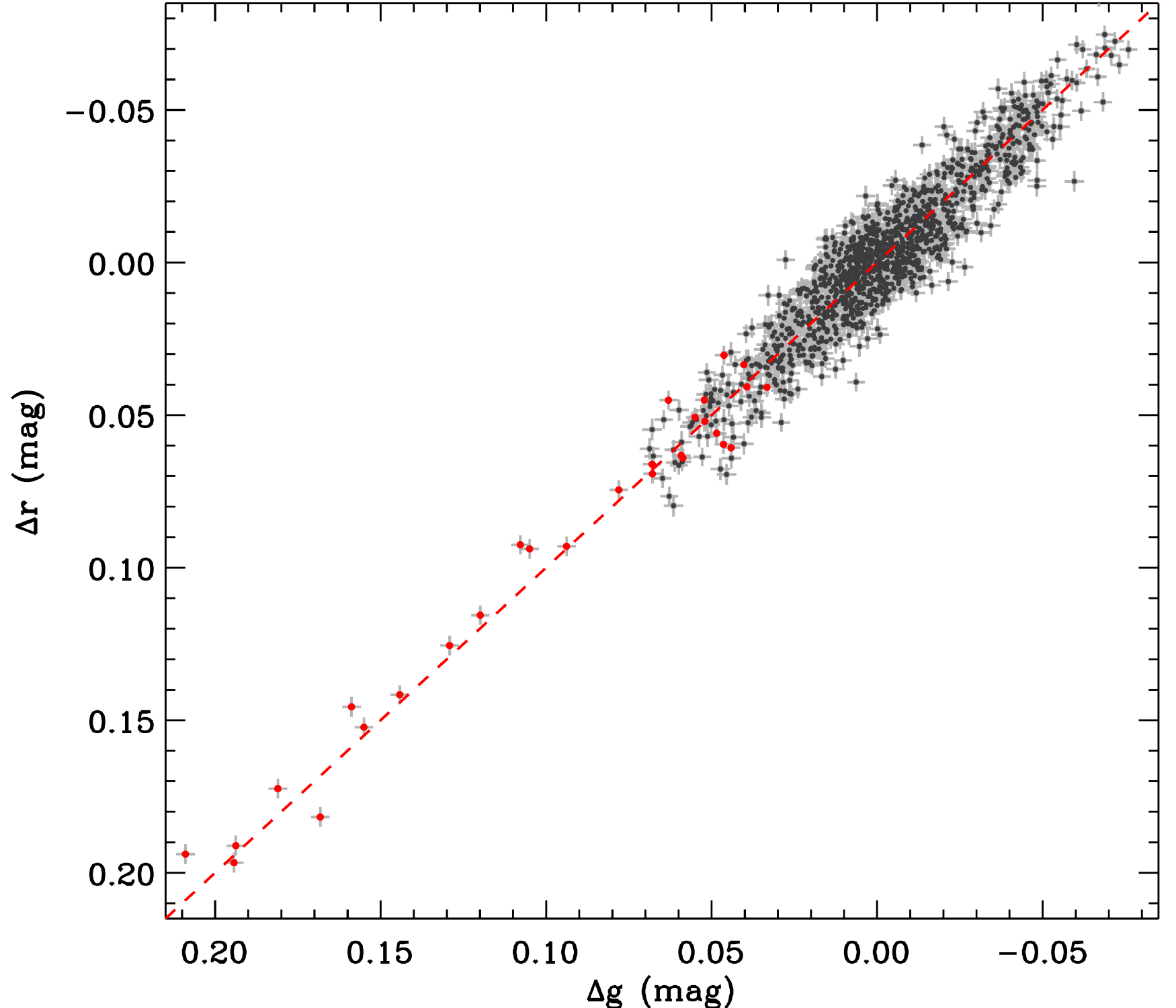}
\caption{Lack of color variations as demonstrated by ULTRACAM data taken on 2019 May 4.  The top panel plots color-coded and labelled $\Delta u$-, $\Delta g$-, and $\Delta r$-band light curves in blue, green, and red, respectively, binned to equivalent time intervals after removal of a five-knot spline fit.  To match the co-added $u$-band data, this process requires a factor of three reduction in the number of $g$- and $r$-band points, and the resulting $u-g$ and $g-r$ colors are plotted in grey and labelled.  The dashed black lines indicate the location of the drifter on this night.  Points falling between these two lines are highlighted in red in the lower panel, which plots $\Delta r$ vs. $\Delta g$, where the resulting slope is consistent with one as delineated by the dashed red line, and thus no color term is present (and similarly for $\Delta u$ vs. $\Delta g$, not shown).
\label{fig:nocold}}
\end{figure}

This analysis was then applied to all nights in a similar manner, with similar results obtained.  For the purposes of display, the data taken 2019 May 4 were analyzed and shown in Figure~\ref{fig:nocold}.  This run was chosen because it is highly representative of the entire ULTRACAM dataset, but also because it contains the first transit event that was discovered to recur on a period clearly different from all other transit events (referred to previously as a drifter).  The drifting transit component has an observed period of around 25.52\,h, which has been determined by visual alignment of the events, as it was only observed on three occasions (2019 May 3, 4, and 5).  While the true period could be shorter, the importance of this drifting transit feature is that is it the deepest transiting event yet witnessed, and thus provides further phase-space coverage in the calculation of the extinction coefficient.

The 2019 May 4 data lead to $\Delta(g-r)/\Delta g=-0.005\pm0.009$ and $\Delta(u-g)/\Delta g=-0.009\pm0.021$, both consistent with zero at high precision.  Not only are these values completely different from what is expected from interstellar extinction -- $\Delta(g-r)/\Delta g=0.287\pm0.005$ and $\Delta(u-g)/\Delta g=0.315\pm0.014$, but they are also inconsistent with a 7900\,K blackbody that varies because of surface temperature fluctuations (0.211 and 0.329, respectively), thus demonstrating that the observed variability cannot be stellar (spots or pulsations).  Data from 2019 March 25 were also of excellent quality and result in coefficients of $\Delta(g-r)/\Delta g=-0.007\pm0.009$ and $\Delta(u-g)/\Delta g=-0.020\pm0.020$. The implications of these grey transits are discussed further in Section~4.3.

%%% FIGURE NO COLOR DETREND %%%

%%% FIGURE UVES VS XS %%%
\begin{figure}
\includegraphics[width=1.0\linewidth]{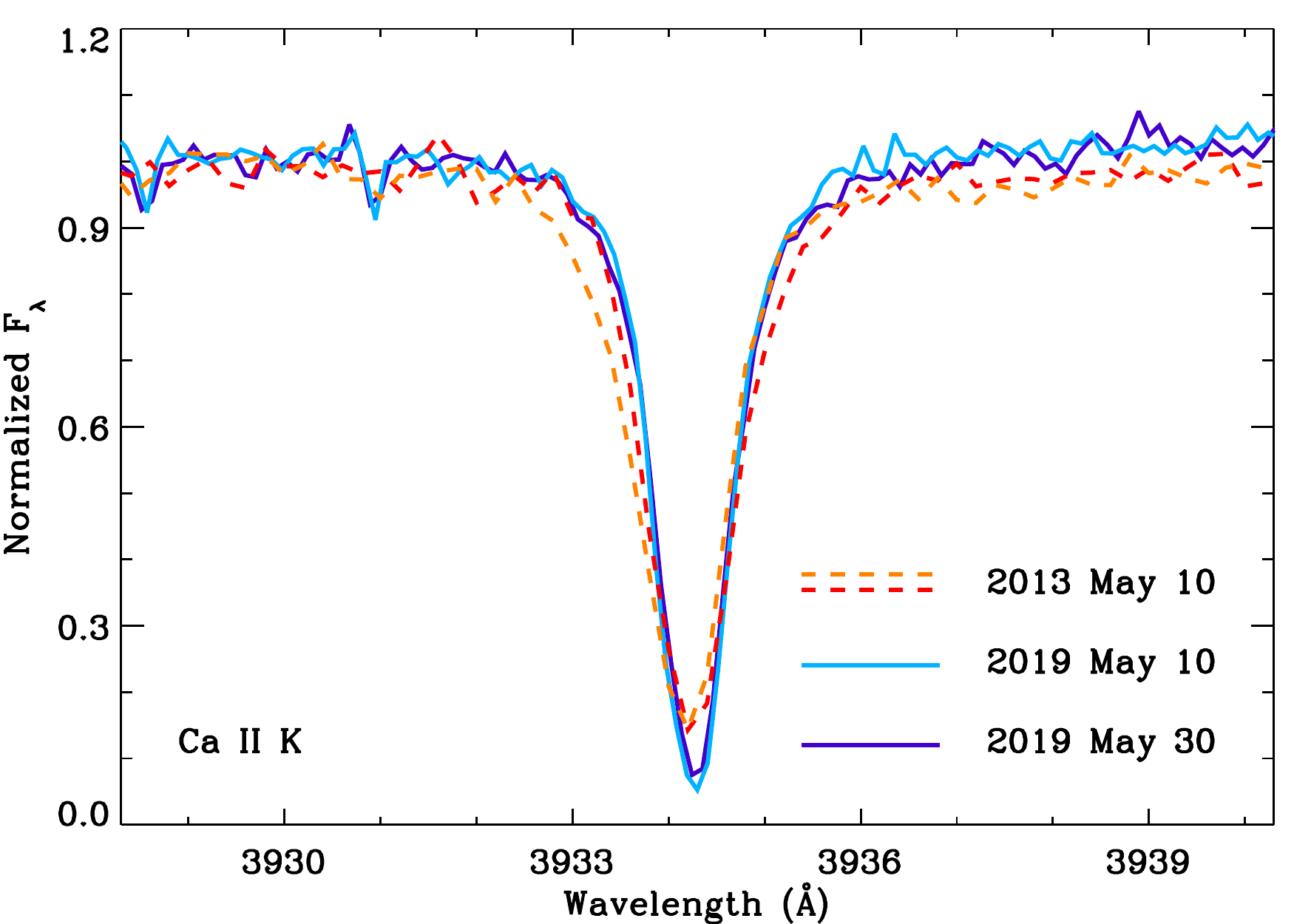}
\caption{The Ca\,{\sc ii}\,K line in WD\,1054--226 as seen during two epochs of UVES spectroscopy in 2019 May, as compared with data taken in 2013 May using X-shooter \citep{vennes2013}.  The resolution of the UVES data have been reduced by a factor of four so that the resolving power of both instruments are similar.
\label{fig:uvesxs}}
\end{figure}

\subsection{UVES: apparent lack of circumstellar gas absorption}

The UVES spectra reveal broad Balmer absorption and narrow, photospheric metal lines.  In Figure~\ref{fig:uvesxs}, the UVES spectra in the Ca\,{\sc ii}\,K line region are compared with medium-resolution (resolving power $R\approx9000$), archival X-shooter spectra taken six years earlier \citep{vennes2013}.  No significant differences can be seen in this line or any of the various metal absorption lines between 2013 May and 2019 May.  The same is true for the UVES data itself; each line detected in multiple (either four or eight) observing runs is unchanging.  It is noteworthy that \citet{vennes2013} inferred the presence of a subtle, circumstellar absorption component in the Ca\,{\sc ii}\,K line, 20\,km\,s$^{-1}$ blue-shifted as compared to predictions of their atmospheric model, but not in any other detected lines of this element or other photospheric metal species.  The UVES data do not confirm this finding despite $4\times$ higher spectral resolution compared to X-shooter. 

\begin{figure*}
\includegraphics[width=1.0\linewidth]{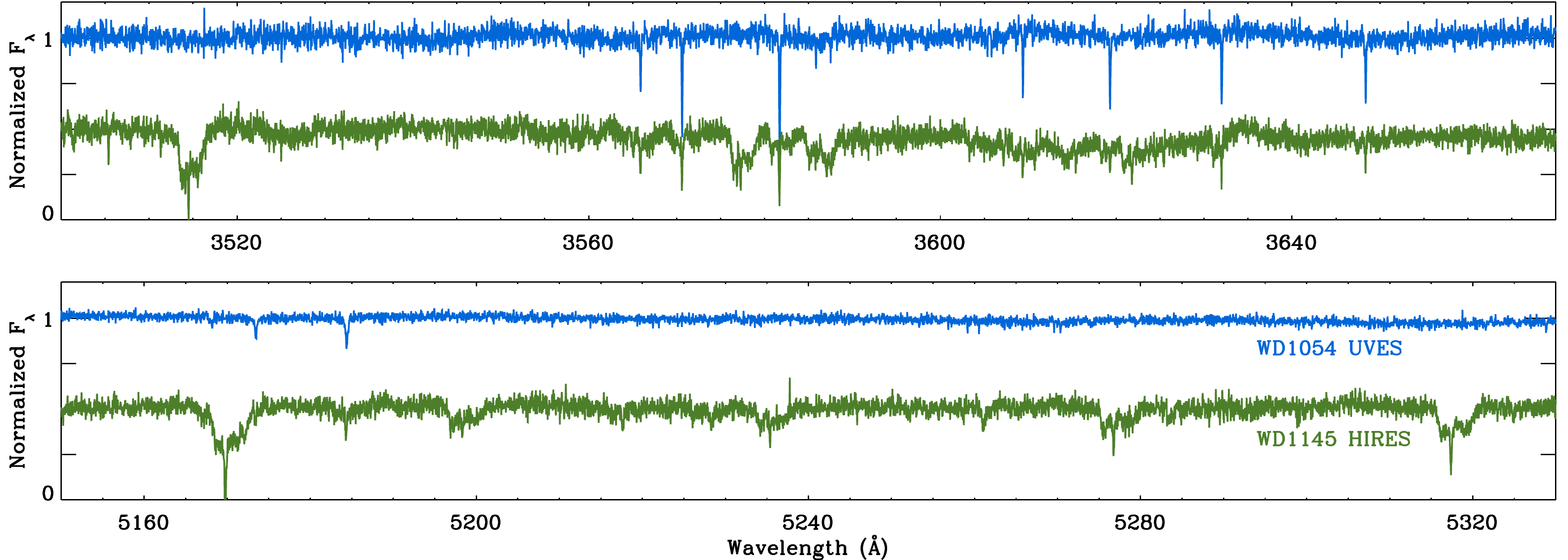}
\caption{Two segments of the UVES spectra for WD\,1054--226 plotted in blue, where narrow, photospheric Mg\,{\sc i} and Fe\,{\sc i} lines can be seen.  In contrast, while many of these same lines are present in the HIRES spectrum of WD\,1145+017 plotted in green, there are numerous Fe\,{\sc ii} lines of circumstellar origin \citep{xu2016,redfield2017}.  Both spectra have similar instrumental resolving power ($R\approx40\,000$) and comparable S/N.
\label{fig:uveshires}}
\end{figure*}

The high-resolution, optical spectral lines of WD\,1054--226 are distinct from the unambiguous circumstellar absorption exhibited by WD\,1145+017 \citep{xu2016,redfield2017}.  Figure~\ref{fig:uveshires} plots the $R\approx40\,000$ spectra of both stars, where WD\,1145+017 has multiple, broad circumstellar features in addition to sharp photospheric components.  In contrast, WD\,1054--226 has only narrow photospheric metal lines.  Further analysis of the UVES data, including the determination of heavy element abundances, will be published separately.

The UVES data provide an upper limit of $v\sin{i}\leq5$\,km\,s$^{-1}$ based on the absence of line broadening by stellar rotation.  A line width analysis was applied to three isolated and relatively strong Fe\,{\sc i} lines (3647.8, 4045.8, and 4404.8\,\AA). These are lines for which photospheric absorption has been detected in cool white dwarfs with metal polluted atmospheres (e.g.\ \citealt{kawka2012a}), and that have relatively weak circumstellar absorption in WD\,1145+017, particularly compared to the Ca\,{\sc ii}\,K line \citep{redfield2017}.  For these reasons, the absorption profiles of these lines in WD\,1054--226 are likely purely photospheric and provide an excellent constraint on stellar rotation.  The resulting upper limit of 5\,km\,s$^{-1}$ implies a lower limit on the stellar rotation period of $p>3.0$\,h, using the parameters from Table~\ref{tbl:sparams}.

\subsection{{\em Spitzer}: lack of infrared excess emission}

\begin{figure}
\includegraphics[width=1.0\linewidth]{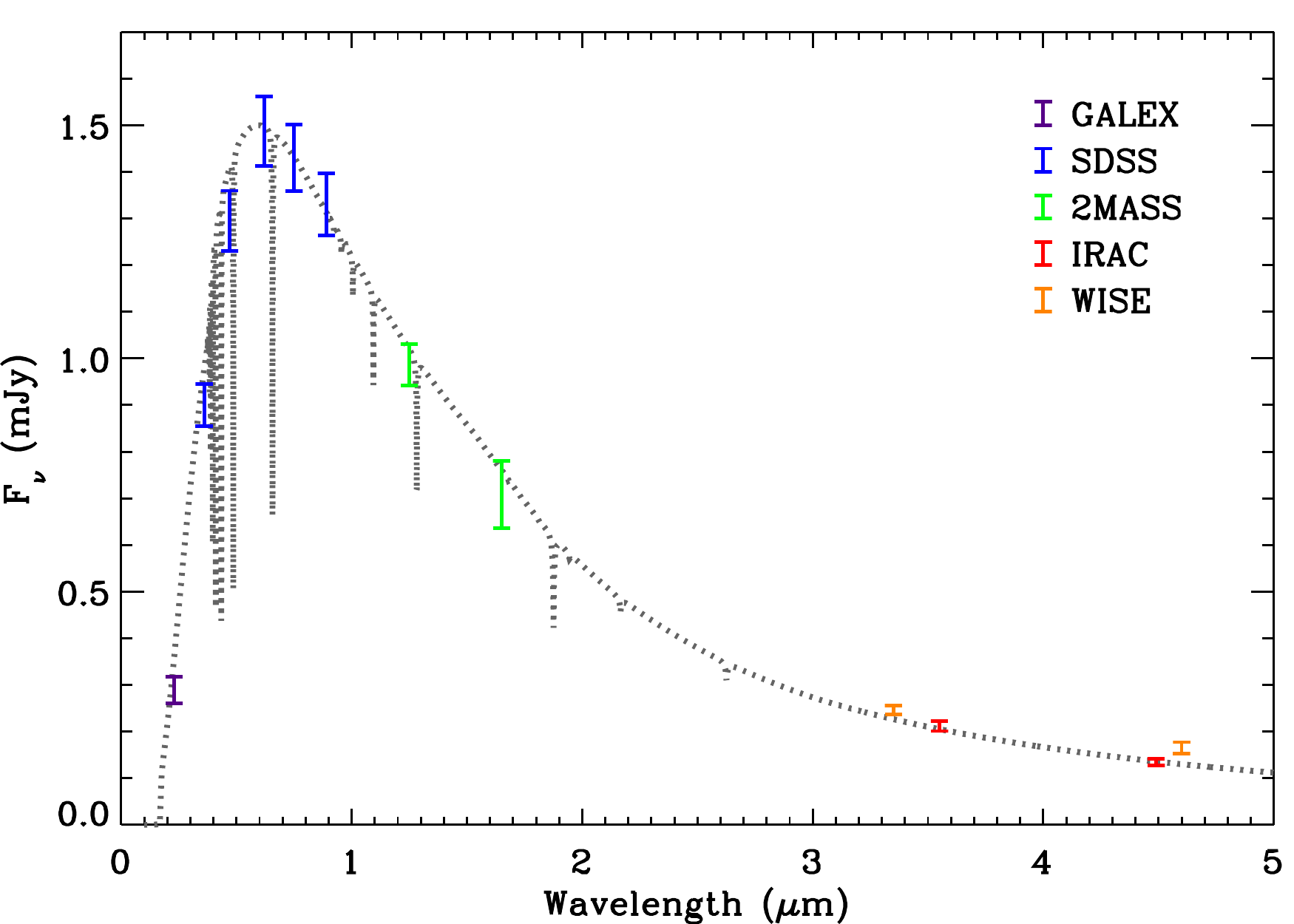}
\caption{Ultraviolet through infrared, photometric spectral energy distribution of WD\,1054--226.  The dotted black line is a hydrogen atmosphere, $T_{\rm eff}=8000$\,K white dwarf model \citep{koester2010}, and the colored error bars are ultraviolet, optical, and infrared photometry with the sources labelled in the plot (and realistic error estimates for {\em Galex} and SDSS where only formal errors are given).  The {\em Spitzer} IRAC measurements are the average of 42 individual 30\,s exposures in each channel, taken consecutively in the 3.6\,$\upmu$m channel 1 and subsequently in the 4.5\,$\upmu$m channel 2.  Across the frame-by-frame exposures in each IRAC channel, there is no evidence for variation more than $3\upsigma$ above the photometric error nor standard deviation of the measurements.
\label{fig:sed}}
\end{figure}

\begin{figure}
\includegraphics[width=1.0\linewidth]{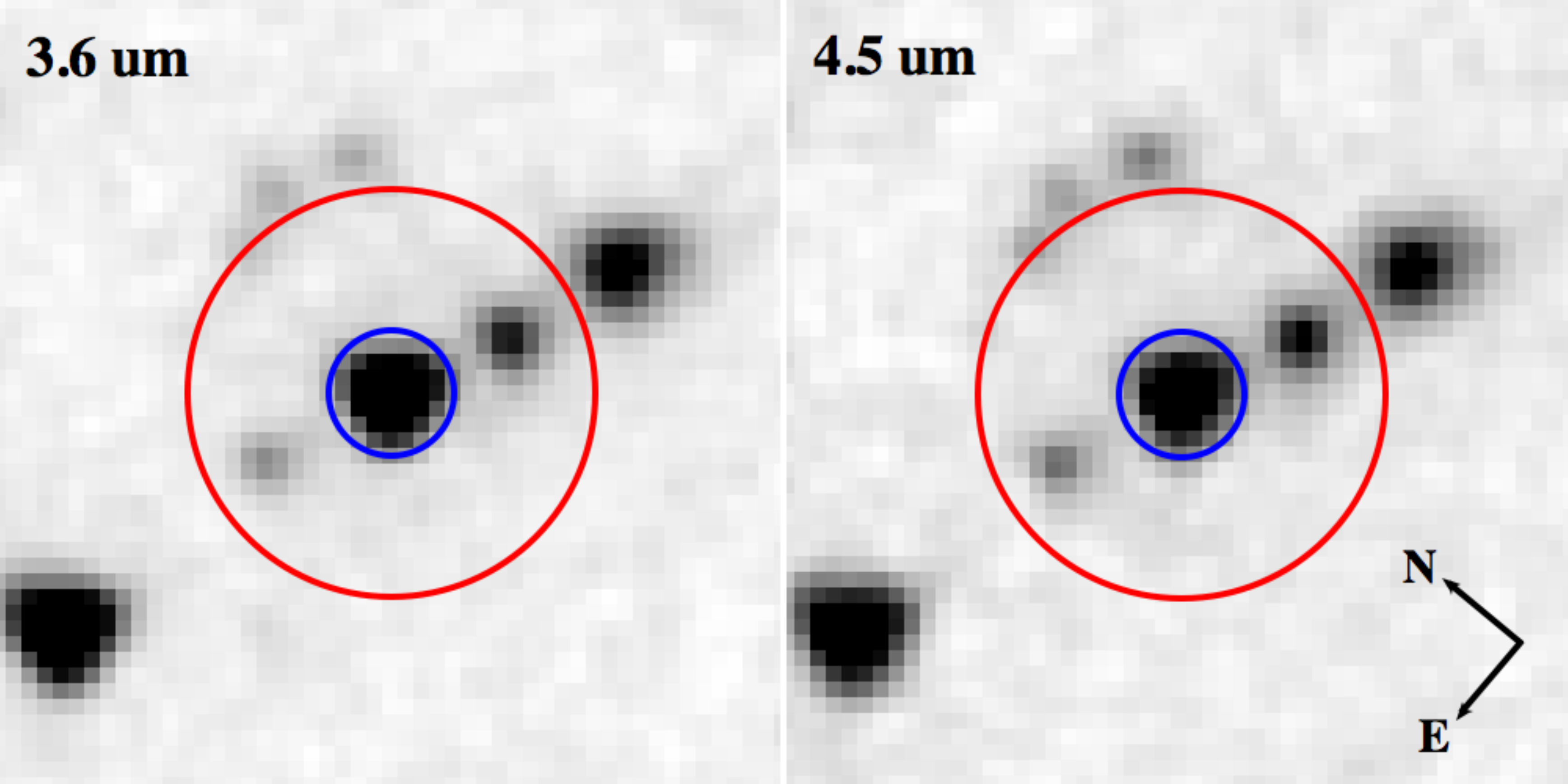}
\caption{Warm {\em Spitzer} IRAC images of WD\,1054--226 with a $30\times30$\,arcsec$^2$ field of view.  The 2.4\,arcsec apertures used for the flux measurements are shown in blue, while a 7.8\,arcsec red circle is shown representing $1.3\times$ the {\em WISE W1} point-spread function that is representative of photometric contamination \citep{dennihy2020} for faint sources such as white dwarfs.  It is thus likely that {\em WISE} photometry for this star suffer from source confusion, and this is consistent with the lower fluxes measured using the IRAC data.
\label{fig:irac}}
\end{figure}

Figure~\ref{fig:sed} displays the multi-wavelength, photometric spectral energy distribution of WD\,1054--226 including the $3-5\,\upmu$m fluxes as measured by {\em Spitzer} and {\em WISE}.  It is important to note that all available, multi-wavelength photometry is likely subject to the intrinsic variability of the dimming events, and thus there is some additional uncertainty in the absolute flux level of the stellar photosphere.  However, the spectroscopically derived $T_{\rm eff}$ and $\log\,g$ used here, should remain unaffected by the grey transits.  The stellar images shown in Figure~\ref{fig:irac} suggest that the {\em WISE} fluxes are contaminated by one or more neighboring sources, while the more reliable {\em Spitzer} fluxes are within $1\upsigma$ (5\,per cent, calibration-limited) of the photospheric model expectations.  Barring the above caveat, the data do not support any infrared excess emission.

%%% FIGURE UVES VS HIRES %%%

\section{Discussion}

This section provides some inferences based on the observed transits and ancillary data for WD\,1054--226, where the data are interpreted as obstruction by clouds of dust or related structures ultimately derived from orbiting planetary bodies.

\subsection{Characteristics of the light curve}

The following is a summary of the WD\,1054--226 light curve:

\begin{enumerate}

\item{The source is highly variable, showing quasi-continuous dimming events of a few to several per cent depth.}

\smallskip
\item{The events occur roughly every 23.1\,min, but not always every 23.1\,min, and are variable in depth, shape, and duration.}

\smallskip
\item{The entire detailed pattern of depth, shape, and duration repeats nearly perfectly every 25.02\,h for at least several orbital cycles.}

\smallskip
\item{The exact morphology of the light curve evolves significantly over longer timescales, but retains significant variability at many integer harmonics of the 25.02\,h orbit, especially the 65$^{\rm th}$ harmonic at 23.1\,min.}

\smallskip
\item{There appears to be at least one, and possibly two additional periodicities, partly evidenced by drifting transit features.}

\smallskip
\item{There is no significant color seen in the $ugri$ bands, to less than 0.02\,mag.}

\smallskip
\item{There is no portion of the light curve that appears consistent with out-of-transit starlight.}

\end{enumerate}

The light curve of WD\,1054--226 is unusual in that the star remains occulted for the entire orbital period of the transiting matter.  This is consistent with the observer looking through a ring of debris, where the occluded fraction of the stellar disk is always non-zero, and, based on the grey transits, consistent with a viewed disk edge that is opaque.  The shapes of the transits are broadly similar to some of the dippers observed towards main-sequence stars where secondary, planetesimal debris structures are suspected to cause the light curve dimming events \citep{gaidos2019,melis2021}, in contrast to the primordial disk structures suspected in the case of many dippers around younger, pre-main sequence stars \citep{ansdell2016,kennedy2017}.  And although the transit shapes of WD\,1054--226 are more complex, and potentially overlapping, there are some features that resemble the exocometary transiting structures seen by {\em Kepler} \citep{rappaport2018b,ansdell2019}.

\subsection{Orbital geometry}

The period of 25.02\,h places the semimajor axis at 3.69\,R$_{\odot}$ for a $M_*=0.62$\,M$_{\odot}$ host star, while for $T_{\rm eff}=7910$\,K, $R_*=0.012$\,R$_{\odot}$, and circular orbits the equilibrium temperature of blackbodies is 323\,K and thus in the habitable zone of WD\,1054--226.  The geometry of the orbit remains uncertain and cannot be directly inferred from the data in hand.  While the circumstellar debris is strikingly clear from the dimming events, there is currently no evidence for the $T\approx1000$\,K dust emission that is commonly seen around polluted white dwarfs (e.g.\ \citealt{xu2020}), where the debris is generally inferred to be located within or near the Roche limit \citep{farihi2016a}.  

\begin{figure}
\includegraphics[width=1.0\linewidth]{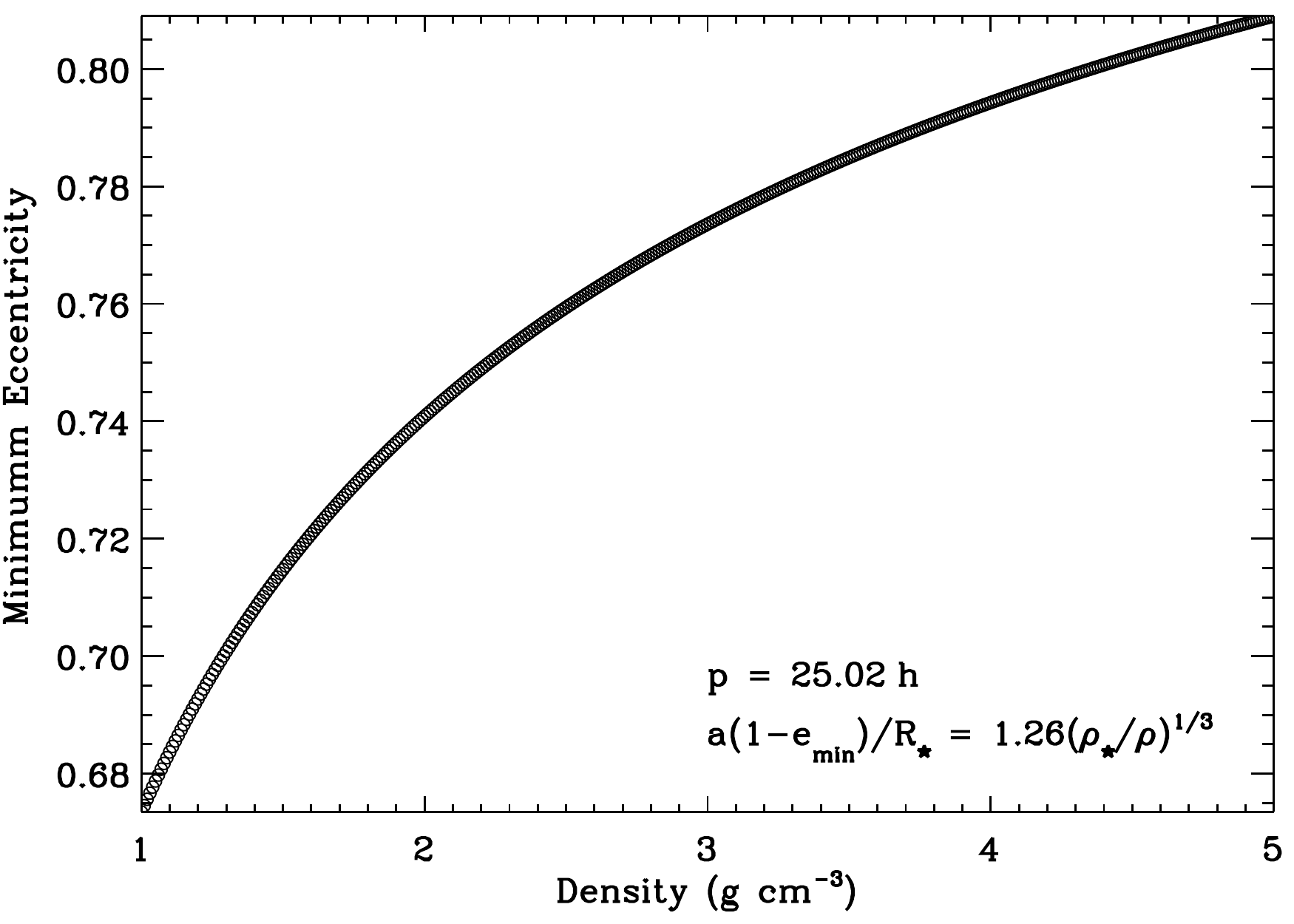}
\caption{Plot demonstrating the minimum eccentricity vs. planetesimal density, for orbits with semimajor axis corresponding to 25.02\,h and periastra within the Roche limit.  The densities plotted are from water ice up to the bulk Earth, and the limit for catastrophic fragmentation is calculated for when self-gravity (no rotation, no material strength) is equal to the tidal force.
\label{fig:eccdens}}
\end{figure}

Two possible orbital geometries are explored, either circular in a disk-like configuration, or highly eccentric with a periastron near or within the Roche limit.  On the one hand, a circular or low eccentricity orbit would imply the underlying parent bodies and their debris are distant from the polluted surface of the star, where ongoing or recent accretion is necessary.  An orbiting ring of debris within the habitable zone would emit at a narrow range of dust temperatures near the equilibrium temperature corresponding to the semimajor axis.  On the other hand, if the observed transiting debris is currently passing within the Roche limit, then the orbital eccentricity is $e\ga0.7$ (Figure~\ref{fig:eccdens}), and there will be signatures of both warmer and cooler dust emission, from material as it orbits near periastron and apastron, respectively.  Such an eccentric orbit would provide a natural explanation for the origin of the debris, because if the underlying bodies, that are the ultimate source of the material causing the transits, are passing sufficiently close to the star to become disrupted by tidal forces, or collide with other planetesimals sharing similar periastra, then this would provide a means for producing numerous, disintegrating fragments \citep{malamud2021,li2021}. For an eccentric orbit, the argument of periastron is unconstrained, and it could be reasoned that the system is most likely viewed close to periastron based on this resulting in the highest transit probability for a dust cloud of fixed size.  However, in the case of a disk with a constant scale height to orbital radius ratio, the transits would be deepest at apastron, providing an argument in favor of this geometry.

Sensitive observations in the infrared with {\em JWST} can distinguish these two possibilities.  This fact is demonstrated in Figure~\ref{fig:detlims}, which plots the sensitivity to single-temperature dust emission at strength $L_{\rm IR}/L_*$ as a function of dust temperature and orbital radius for WD\,1054--226.  Using the relationship between the fractional dust luminosity, the level of dimming, and orbital radius during transit \citep{wyatt2018}, 323\,K dust causing 5\,per cent dimming should have $L_{\rm IR}/L_*\approx6\times10^{-4}$ if all the debris is within a narrow annulus in the habitable zone.  However, if the dust is optically thick and observed edge-on, then the disk will appear fainter because emission will be largely obscured within the plane of the disk.  Moreover, Section~4.3 suggests that the material is indeed likely optically thick, and must be viewed edge-on to cause transits.  Nevertheless, Figure~\ref{fig:detlims} shows that the optically thin case is achievable with a relatively modest integration time in at least two of the longer wavelength filters of MIRI.  

For the eccentric orbit case, the precise periastron distance has implications for the subsequent evolution of debris generated initially via tidal disruption.  Specifically, the potential production of gaseous debris, via sublimation or collisions, can assist in disk compaction \citep{malamud2021}, provide material for surface pollution on viscous timescales \citep{farihi2012b}, thus providing a means of dust removal, or possibly re-condense into solids that comprise the opaque transiting clouds \citep{metzger2012}.  In the case that periastron corresponds to the Roche limit for material of density 3\,g\,cm$^{-3}$, the corresponding blackbody temperature is just under 680\,K, and no refractory planetary material would effectively sublimate under these conditions.  Therefore, unless the underlying parent bodies of the transiting structures are composed of water or similarly volatile ices -- which is not reflected in the photospheric metal pollution -- it is likely that any ongoing dust production and planetesimal disintegration are driven by collisions.  While 1400\,K is often cited as a temperature where silicates are rapidly sublimated in e.g.\ hydrogen-rich protoplanetary disks, it has been shown that for volatile-poor debris orbiting white dwarfs, there should be high metal gas pressure (no hydrogen) and a typical benchmark temperature that is 300--400\,K higher \citep{rafikov2012}.  In this case, periastron would need to lie at roughly 0.1\,R$_{\odot}$, requiring $e>0.96$ and is thus dynamically less likely.

%%% FIGURE SED %%%

%%% FIGURE IRAC IMAGE %%%

\subsection{Occulting clouds}

There is no evidence for solid-body transits in any of the data, and the underlying transit sources are presumably relatively small but with extended comae of dust and gas.  It cannot yet be determined what fraction of the stellar disk is constantly occluded, which could be significant.  But taking a typical transit depth for WD\,1054--226 as 3\,per cent, this corresponds to a spherical blocking radius of around 1480\,km, and would be between the mean radii of solar system moons Triton and Europa.  However, the duration of the transit events provides clear indications that the occulting clouds are non-spherical.  

Transit durations range from narrow events lasting $4-8$\,min, to broader dips whose widths are $12-15$\,min.  Together with the Keplerian speed for a circular, 25.02\,h orbit, this implies the clouds are vastly spread along the orbit, on the order of several $10^4$\,km up to $10^5$\,km, and typically an order of magnitude larger than the stellar radius.  This implies that the scale height of the optically thick clouds is better approximated assuming a rectangular shape, where at any given time the entire stellar diameter is occluded along the orbit.  For an eccentric orbit of $e\approx0.7$, the above approximation needs modification for the varying orbital speed, where if viewed at periastron or apastron the implied length of a typical transiting cloud would be roughly twice as long or twice as short, respectively.  Using this crude shape estimate for the transit geometry and a circular orbit, a typical 3\,per cent transit depth results in a blocking height of 400\,km along the stellar equator (and thus larger for higher impact parameters).  For comparison, the stable arcs in the Adams ring of Neptune are $10^3$\,km up to $10^4$\,km in arc length \citep{porco1991}, but are unlikely to have a vertical height significantly larger than the ring particles \citep{esposito2002}.

It is tempting to try and estimate the total fraction of the stellar disk that is constantly occluded, but this is not possible even from the sensitive light curves presented here.  This fraction is at least a few per cent, based on the fact that a typical ULTRACAM light curve has maximum flux values that are a few per cent above the normalized, nightly mean.  Another method would be to compare the predicted absolute brightness of the star with that observed, for example based on the spectroscopically derived $T_{\rm eff}$ and $\log\,g$ from model fits to Balmer lines (unaffected by the grey transits) versus those calculated from photometry and parallax.  But it is important to note that such comparisons are well known to be fraught with systematic offsets in derived parameters (e.g.\ \citealt{bergeron2001,genest2014,tremblay2019}), and thus must be viewed with caution.  That being said, comparing the luminosity predicted by spectroscopy \citep{gianninas2011} and that observed by {\em Gaia} photometry and parallax \citep{gentile2019}, results in an observed luminosity that is 10\,per cent fainter than predicted.  Numerous stars in this range of $T_{\rm eff}$ and $\log\,g$ show similar or even larger deviations (in both directions), and therefore any difference in WD\,1054--226 may be indicative, yet far from conclusive.

The lack of color, as determined from the ULTRACAM data, supports optically thick dust clouds or large particles, similar to findings for WD\,1145+017 \citep{alonso2016,izquierdo2018}.  However, there is no efficient way to remove small particles around a white dwarf as cool as WD\,1054--226, and thus an optically thick configuration is more likely, and thus the underlying dust mass is unconstrained by transit observations.  Detection of dust emission in the infrared would place important constraints on the properties of the circumstellar material.  For example, the range of detected emission temperatures would yield the periastron and apastron distances of the dust distribution, and the shape of the emission would indicate the optical depth and possibly the disk orientation \citep{wyatt2018}.  Such observations would thus have broad implications for the overall population of polluted white dwarfs, as WD\,1054--226 is a rather ordinary member of this class, and the evolution of its circumstellar disk is likely relevant to the entire population.

The lack of detectable circumstellar gas is still somewhat surprising, as the scale height of gas is expected to be larger than that of dust (or the same if physically coupled).  However, if the configuration is simply opaque to the line of sight, then no conclusions can be drawn from the non-detection.  In the case of WD\,1145+017, it appears that circumstellar gas absorption is reduced during transits, leading to an increase in blue-wavelength fluxes where a multitude of lines are seen \citep{hallakoun2017,karjalainen2019}.  This is consistent with opaque transiting clouds that (partly) block the line of sight through the circumstellar gas \citep{xu2019a}.  

In WD\,1054--226, one might speculate that the analogous transiting bodies block all such material, and this would be consistent with the utter lack of out-of-transit starlight, whereas in contrast, WD\,1145+017 seems to have out-of-transit flux observed in its light curve as evidenced by relatively flat regions \citep{rappaport2016,gansicke2016}.  It is noteworthy that ZTF\,0139 and ZTF\,0328 may have circumstellar features, although the evidence is currently less compelling \citep{vanderbosch2020,vanderbosch2021} than for WD\,1145+017.  If true, WD\,1054--226 would be unique among transiting systems for its lack of circumstellar absorption.

%%% FIGURE ECC V DENS %%%

\subsection{Non-canonical disk orbiting a typical polluted white dwarf}

The IRAC data are consistent with stellar emission only, and provide the first clear case where a debris disk is not detected via infrared photometric excess, yet whose presence is revealed by the transiting events.  There also does not appear to be any significant infrared excess towards either ZTF\,1039 or ZTF\,0328, although the data available for those stars are not comparable to the sensitivity of the {\em Spitzer} images analyzed here.  All four white dwarfs with irregular transits and optical spectroscopy display metal absorption lines, and thus their surfaces are externally polluted by planetary material.

Among metal-rich white dwarfs, WD\,1054--226 does not stand out in mass, temperature, detected element species or abundances.  Its lack of infrared excess at warm {\em Spitzer} and {\em WISE} is also unremarkable, as the vast majority of isolated and metal-lined white dwarfs exhibit only photospheric emission at wavelengths of 3--4\,$\upmu$m \citep{rocchetto2015}.  However, there are some differences in the disk-detection statistics between warmer (younger) and cooler (older) polluted white dwarfs, albeit with some observational bias that results primarily from the spectroscopic sensitivity to photospheric metal abundance (and hence the implied accretion rate of circumstellar matter).

%%% FIGURE DETECTION LIMITS %%%
\begin{figure*}
\includegraphics[width=1.0\linewidth]{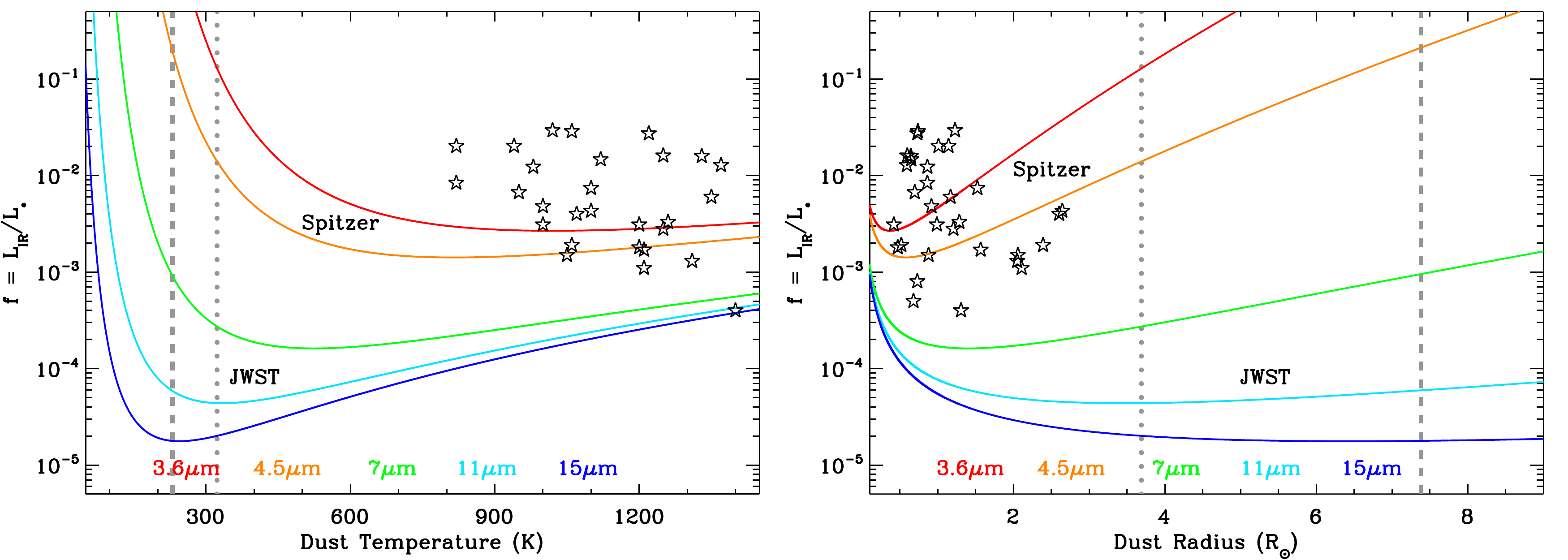}
\caption{The sensitivity to single-temperature dust emission at strength $L_{\rm IR}/L_*$ as a function of dust temperature (left panel) and orbital radius (right panel) for WD\,1054--226.  The red and orange curves show the parameters of dust disks that would have been detected with {\em Spitzer} IRAC at 3.6 and 4.5\,$\upmu$m for this particular host star, while the green and blue curves trace the cool dust sensitivity with {\em JWST} MIRI at 7, 11, and 15\,$\upmu$m (using 18.5\,min of total integration per bandpass).  The grey dotted lines are the expected dust temperature (323\,K) and circular orbital radius in the habitable zone.  The grey dashed line is the expected dust temperature (230\,K) and corresponding orbital radius (roughly twice the semimajor axis) for cooler emission from an eccentric orbit \citep{wyatt2010}; in this case, warmer emission from dust near periastron is also expected.  The 32 star symbols are white dwarf dust disks with well determined $L_{\rm IR}/L_*$ \citep{rocchetto2015}.  Note that some of the plotted symbols in the right panel cannot be seen in the left panel due to higher dust temperatures that place them beyond the plot range.  Furthermore, the colored curves are specifically calculated for WD\,1054--226 whose luminosity is significantly lower than the host stars with plotted dust $L_{\rm IR}/L_*$ values.  Because the conversion from dust temperature to orbital radius depends on the actual host $L_*$ and $T_{\rm eff}$, the plotted symbols move relative to the lines of detectability for WD\,1054--226.
\label{fig:detlims}}
\end{figure*}

For hydrogen-rich white dwarfs with sinking timescales of days or weeks at most ($T_{\rm eff}\ga14\,000$\,K), and sensitive ultraviolet detections of photospheric metals, around 10\,per cent also display infrared excess from their circumstellar disks.  The majority remain undetectable (e.g.\ with {\em Spitzer}), but some circumstellar material must be present based on the requirement for ongoing accretion \citep{wilson2019}.  For similar infrared observations of those stars with ground-based, optical detections of metals, the fraction with observed dust emission is significantly higher, and ranges from 20\,per cent to as high as 50\,per cent, depending on the temperature and age range \citep{farihi2009a}.  This contrast is understood to be a bias favoring infrared dust detection towards stars with higher metal abundances (and accretion rates) that can be more easily detected via optical spectroscopy \citep{kilic2007,girven2012}.  The same is not true for metal-rich white dwarfs with $T_{\rm eff} \la 10\,000$\,K, including WD\,1054--226 \citep{farihi2016a,bonsor2017}.  There are dozens of relatively nearby and bright DAZ and DZ stars in this temperature and age range known, including many with metal abundances (and time-averaged accretion rates) similar to their warmer counterparts.  Nevertheless, only two dusty and polluted stars have been unambiguously identified among those stars with reliable infrared data, and represent a fraction smaller than 4\,per cent \citep{farihi2008b,bergfors2014,debes2019,lai2021}.  

If WD\,1054--226 is representative of polluted white dwarfs with $T_{\rm eff} \la 10\,000$\,K, then it suggests that this population harbors disks, but they are not readily detectable in the infrared for exactly the reasons illustrated in Figure~\ref{fig:detlims}.  The orbital period clearly shows that the material is, at least most of the time, more distant than the $1-2$\,R$_{\odot}$ found for stars with dust emission detected with {\em Spitzer} or similarly {\em WISE} at $3-4$\,$\upmu$m.  A search for cooler dust around polluted white dwarfs, at more distant orbital radii, using sensitive infrared instruments, appears valuable on the basis of WD\,1054--226.

It should be noted that at least 30 polluted white dwarfs were targeted using the MIPS 24\,$\upmu$m camera aboard {\em Spitzer} during the cryogenic mission.  The nearest and brightest DZ white dwarf, vMa\,2, was found to have a 24\,$\upmu$m flux consistent with photospheric emission, but otherwise only those stars with infrared excesses at IRAC wavelengths were detected with MIPS \citep{farihi2009a}.  Only two polluted white dwarfs have published observations at longer wavelengths, G29-38 and GD\,362 \citep{xu2013,farihi2014}, both observed with {\em Herschel} PACS and the former observed also with ALMA, and all resulting in upper limits.  A rigorous analysis of non-detections at wavelengths longer than $3-4$\,$\upmu$m has not yet been investigated \citep{bonsor2017}.  Together with results for the warmer stars where ongoing accretion from circumstellar matter is required, the transit detections support a picture where photospheric metal pollution is nearly equivalent to the detection of extant circumstellar debris \citep{farihi2013b}.

The overall data indicate that the transiting material has significant scale height of at least a few hundred km, and is thus not in a canonical disk configuration that is vertically opaque yet thin \citep{jura2003}.  Such a dynamically relaxed disk may be favored, at least in part, for the most closely orbiting debris disks where, e.g., baseline dust emission is always present despite significant infrared variability observed on timescales of months to over a decade \citep{swan2020}, and the inferred, ongoing accretion rates are well matched by evolutionary models of irradiated flat disks \citep{rafikov2011a,farihi2016a}.  As in the rings of Saturn, flat disk constituents share the same inclination to the observer, and the scale height is comparable to the largest particles.  

However, large bodies can stir up an otherwise-flat disk, to a height that has been found to extend nearly to a Hill radius in the rings of Saturn \citep{hoffman2015}.  Using this approximation, to achieve a scale height of 400\,km towards WD\,1054--226, the largest bodies would need to be $\sim50$\,km in radius for typical asteroid densities.  This calculation is based on a circular orbit, but does not change drastically for the eccentric case, where the instantaneous Hill radius scales as orbital distance.  In the extreme case of high eccentricity at apastron, a 400\,km Hill radius corresponds to a body of radius $\sim25$\,km.  It is noteworthy that 400\,km is an order of magnitude smaller than some predictions of a collisional cascade in a narrow annulus around white dwarf \citep{kenyon2017a}, where analogous processes set the scale height.

If transit activity seen towards white dwarfs were caused by an order of magnitude or more increase in a typical disk scale height, due to dynamically activated dust production (i.e.\ tidal disruption, collisions, outgassing), then those debris disks would become more amenable to infrared detection \citep{kenyon2017a,bonsor2017} regardless of disk inclination.  While transits are necessarily observed near to edge-on, the surface area presented by the substantial dimming events can be sufficient to result in detectable dust emission \citep{vanderburg2015,wyatt2018}, but the actual orbital radius during transit (temperature), and optical depth, are also important.  The data are insufficient to form any conclusions at present, but given that only one of four well-studied transiting systems has an infrared excess implies that the transits result from favorable geometry, and the scale height of the occulting clouds are typical (and undetectable in the infrared for both transiting and non-transiting systems).

\subsection{Origin of regular dimming events}

The origin of the 23.1\,min period is puzzling.  It may be notable that the temperature of circumstellar material with such an orbital period would be close to 1300\,K.  This is $300-400$\,K cooler than where the bulk of planetary solids would sublimate in the presence of metal-dominated vapor pressure \citep{rafikov2012}, but certainly where volatiles would be rapidly lost to the gas phase and be prone to viscous accretion onto the star. 

This suggests the possibility of a connection between two spatially distinct regions, where it is possible to imagine a close orbit affecting material further out.  For example, \citet{manser2019} identified a clump near the inner edge of a white dwarf debris disk that is extended azimuthally to roughly 0.4 of the orbit at that location.  While that clump is slightly further out and hotter than that implied by the 23.1\,min period here, that paper describes how clumps could form from vortices at the inner edge of white dwarf disks, and such clumps could lead to variable illumination of the outer disk.  However, the exact commensurability between the 23.1\,min period and that at 25.02\,h argues the dimming events have a more distant origin.

Another possibility that was ruled out is that the 23.1\,min scale transits arise from individual fragments created in the break-up of a large parent body, all of which remain on a 25.02\,h orbit.  The 23.1\,min periodicity would then correspond to the regular spacing of the resulting fragments, motivated by the 'string of pearls' seen in the fragmented comet Shoemaker-Levy 9 \citep{weaver1994}.  However, while the fragments of a tidal disruption might be expected to be separated by some characteristic radius (set by the velocity imparted to the fragments), they would neither be expected to be exactly regular, nor to extend to so many clumps all yielding similar transit depth.

A period of 23.1\,min could arise naturally within material orbiting at 25.02\,h period if 23.1\,min is the difference between two similar orbital periods. This could arise from mean motion resonances in which the material causing the transits orbits the star 65 times for every 66 orbits of an unseen perturber.  Such resonances are implicated in a variety of structures seen in planetary rings, where the perturbers are moons.  For example, eccentricity (radial) or inclination (bending) waves in a disk, such as those seen in the rings of Saturn, can be relatively regularly spaced.  However, these are also critically damped in both the leading and trailing directions from the perturber \citep{borderies1984,porco1987,weiss2009}, and there is no evidence of such in the light curve of WD\,1054--226.  Furthermore, the light curve dimming is clearly not sinusoidal as are the models of undamped edge waves in planetary rings. 

An external, 66:65 mean-motion resonance with a planet orbiting at 25.02\,h is plausible, and would result in an exact commensurability between individual resonant structures and period as observed.  In this case, the entire pattern of clumps rotates around the star, following the planet, at sub-Keplerian speeds \citep{wyatt2003}.  For such a resonance to be stable, i.e.\ to fall outside the resonance overlap criterion established by \citet{wisdom1980}, two cases are considered.  For a collisionless n-body, the unstable region around a perturber is given by the width of the chaotic zone of overlapping resonances \citep{duncan1989}, and yields a mass of 0.0055\,M$_{\oplus}$ if the 66:65 resonance is the last that is stable.  It has been predicted that such small planets can form close to a white dwarf, following the tidal disruption of a sufficiently massive body such that viscous disk evolution dominates \citep{vanlieshout2018}, although these do not migrate outward to account for the current observations.

Using a fluid approximation instead, assuming the collisional disk viscosity replenishes particles in the chaotic zone, material will spread to up to the Hill sphere of the perturber (similar to opening a gap in a gaseous protoplanetary disk).  If the 66:65 mean-motion resonance lies at the edge of the Hill sphere, this results in a significantly higher perturber mass of 0.66\,M$_{\oplus}$.  However, in either case, while the resulting arrangement of clumps may be highly regular in such a resonance, this interpretation would require individual particles to move between clumps, and therefore the observer would see different material at subsequent apparitions of a recurring transit event.  In contrast, e.g.\ the double-dip feature observed to repeat on several nights toward WD\,1054--226 has been interpreted to be the same occulting material, and similarly for essentially all other transit features.

One possibility that has been invoked in planetary rings to account for the lack of radial or azimuthal spread in ring arcs and clumps is resonant confinement.  This has been invoked to account for the stable ring arcs of Neptune, via the eccentricity of its moon Galatea \citep{namouni2002}, and it is speculated that Metis may confine km-sized particulate clumps found in the rings of Jupiter \citep{showalter2007}.  The transiting events seen towards WD\,1054--226 have significant arc length and may more than superficially resemble planetary ring structures such as those observed in the Adams ring.  Broadly speaking, planetary ring resonant formations require perturbations by a body that has comparable mass to the rings they perturb, opening up the possibility that a major planetary body orbits within the habitable zone of WD\,1054--226.

All the above (inadequate) scenarios assume a circular orbital geometry, but neither are any solutions apparent for an eccentric ring.  At present, the exact cause of the overarching transit sub-structure remains unknown.

\subsection{Drifting transits}

It is noteworthy that the 25.5\,h period, as traced by one of the drifting transit components, may be similar to the orbit of the primary transits (though shorter-period aliases remain possible).  This also appears to be the case for WD\,1145+017 \citep{gary2017} and ZTF\,0328 \citep{vanderbosch2021}.  In WD\,1145+017, it appears that all the drifting transit features orbit more rapidly than the fundamental period and are thus closer to the star, leading to an interpretation that these more rapidly orbiting structures are due to tidal disruption at the L1 point \citep{rappaport2016}.  This may not be the case for WD\,1054--226, where the deepest transiting event appears to recur more slowly and is potentially located further from the star.  In ZTF\,0328 it is not clear if either of the two observed periodicities is dominant, but their similarity is unlikely to be coincidental.  The overall picture emerging from transiting white dwarfs appears consistent with a single, dynamically related family of orbital elements in each case.  

The number of underlying parent bodies on the 25.02\,h period of WD\,1054--226 is unknown; they may be legion, waves or resonant structures in a debris disk, or dust clouds held in resonant confinement.  Regardless, there is as yet no evidence for any single parent body that is undergoing disintegration. 

In both WD\,1054--226 and WD\,1145+017, the data are consistent with optically thick clouds of debris, and it should be remarked that these would potentially occult any shorter-period transits with identical inclinations.  The 25.5\,h drifting transit component seen towards WD\,1054--226 does not appear to block the 25.02\,h signal where it occurs in the light curves.  This inference comes from the co-added light curve, which is not adversely affected in the relevant phases where the slower drifter is present.  But more acutely, if the 25.02\,h period transits are optically thick, the 11.4 (22.9)\,h transiting bodies are likely to have a different inclination (i.e.\ both corresponding signals are detected in {\em TESS}).

For WD\,1054--226, in order to simultaneously view the full height (e.g.\ 400\,km) of occulting, optically thick clouds at different orbital radii requires that a flat disk is inclined towards the observer, whereas the transits themselves require a nearly edge-on view.  While the uncertain orbital geometry and line of sight preclude further numerical constraints, it is unlikely that the transiting components at different periods all share a single inclination.  This is consistent with a process that alters the orbits of transiting bodies, and underscores a dynamical environment.  As WD\,1054--226 shares this property with at least WD\,1145+017, it is tempting to identify slightly different periods with slightly different inclinations in transiting white dwarf systems.

It may be the case that planetary debris is generated from the interaction of numerous fragments from an initial tidal disruption, where ensuing collisions are primarily responsible for the evolution \citep{kenyon2017a,malamud2020,malamud2021}, and perhaps the observed transits with different periodicities.  It is noteworthy that the main Jovian ring is thought to be generated by sufficiently continuous impact events on the two ring moons Metis and Adrastea \citep{depater2018}, where trailing clumps of debris have been observed by {\em New Horizons}, and possibly held by resonant confinement \citep{showalter2007}.  Similar impacts could be ongoing for analogous parent bodies orbiting white dwarfs, if the environment is sufficiently rich in potential impactors.

\section{Outlook}

If one assumes that 25\,h is crudely representative of the orbital periods where transits might be detected, then the transit probability for a typical white dwarf is $R_\star/a\approx0.4$\,per cent.  If the orbital geometry passes through the Roche limit and so $e\ga0.7$ then this increases to $R_\star/a(1-e)\ga1.2$\,per cent for transits seen at periastron \citep{barnes2007}.  There are 1048 high-probability white dwarfs within 40\,pc \citep{gentile2019}, including WD\,1054--226, and it is well-established that at least 30\,per cent of white dwarfs are polluted with photospheric metals that originate as circumstellar, planetary material.  While highly uncertain, this would imply that there should be $1048\times0.3\times0.004-0.012\approx1-4$ transiting and polluted white dwarfs within the local 40\,pc volume, if the debris surrounding WD\,1054--226 is representative of all polluted white dwarfs.  It is notable that ZTF\,0328 lies not far outside this volume, but it seems plausible that there are other local volume examples to be found, especially for smaller semimajor axes and higher eccentricities.  A statistically robust lack of transits, at a given frequency for polluted white dwarfs, would imply that the source regions of the parent bodies responsible come from relatively distant orbits.

The transiting events detailed here for WD\,1054--226 are unprecedented in detail and coverage, and paint a picture of material that is constantly occluding the star, and at least partly orbiting outside the Roche limit (and in this sense similar to ZTF\,1039 and ZTF\,0328).  From a dynamical and evolutionary perspective, the origin of the large occulting clouds is likely the result of tidal disruption or collisions in the vicinity of the Roche limit, where there is a narrow range of periastra for the material.  However, other possibilities cannot be ruled out, and {\em JWST} is necessary to detect and characterize the debris disk.

Prior work has made it clear that circumstellar disks are often (or always) present around polluted white dwarfs, via the need for ongoing accretion of circumstellar matter to account for photospheric metals.  And because the infrared detection of this orbiting material is typically curbed by sensitivity limits, the transits presented here provide additional insights that are not available otherwise. Thus, WD\,1054--226 provides potential evidence of a typical white dwarf debris disk.  That is, these data are consistent with previous work that indicates typical infrared dust emission is significantly less luminous than those detected to date, well beyond the sensitivity of {\em WISE} and even {\em Spitzer}, but amenable to longer wavelength observations.

\section*{Acknowledgements}
The authors thank the reviewer S.~Rappaport for an enthusiastic and careful report.  The team acknowledges the European Southern Observatory for the award of telescope time via programs 0102.C-0700 and 0103.C-0247 (PI Farihi), and the La Silla Observatory for the support to carry out the ULTRACAM observations.  The UVES spectra were obtained via DDT award 2103.C-5013, and the {\em Spitzer} observations were carried out under a time-critical DDT program 14261 (PI Farihi).  This paper includes data collected by the {\em TESS} mission, whose funding is provided by the NASA Explorer Program.  This work has made use of the ESA {\em Gaia} mission, processed by the {\em Gaia} DPAC.  J.~Farihi thanks C.~J.~Nixon and J.~E.~Pringle for useful discussions, and acknowledges support from STFC grant ST/R000476/1.  J.~J.~Hermes acknowledges support through {\em TESS} Guest Investigator Program 80NSSC20K0592.  T.~R.~Marsh acknowledges the support of STFC grant ST/T000406/1, and from a Leverhulme Research Fellowship.  A.~J.~Mustill acknowledges support from the Swedish Research Council (starting grant 2017--04945).  T.~G.~Wilson acknowledges support from STFC consolidated grant number ST/M001296/1.  A.~Aungwerojwit and C.~Kaewmanee received support from Thailand Science Research and Innovation grant FRB640025 contract no.\ R2564B006.  V.~S.~Dhillon and ULTRACAM are supported by the STFC.

\section*{Data Availability}
Data acquired at ESO facilities for the PI programs listed above are available through their archive, with the exception of ULTRACAM data, which are available on request to the instrument team.  Space-based data from {\em Spitzer}, {\em TESS}, and {\em WISE} are available through the relevant NASA archives.

\bibliographystyle{mnras}

%\bibliography{references}
\bibliography{/Users/jfarihi/papers/references}

\appendix

\section{Supplementary Light Curve Data and Figures}

For completeness and reader interest, the full set of ULTRACAM light curves are shown here in the Appendix, and including data for individual bandpasses for the first and last nights.

%% FIGURE FIRST + LAST NIGHT IN COLOR %%%
\begin{figure*}
\includegraphics[width=1.0\linewidth]{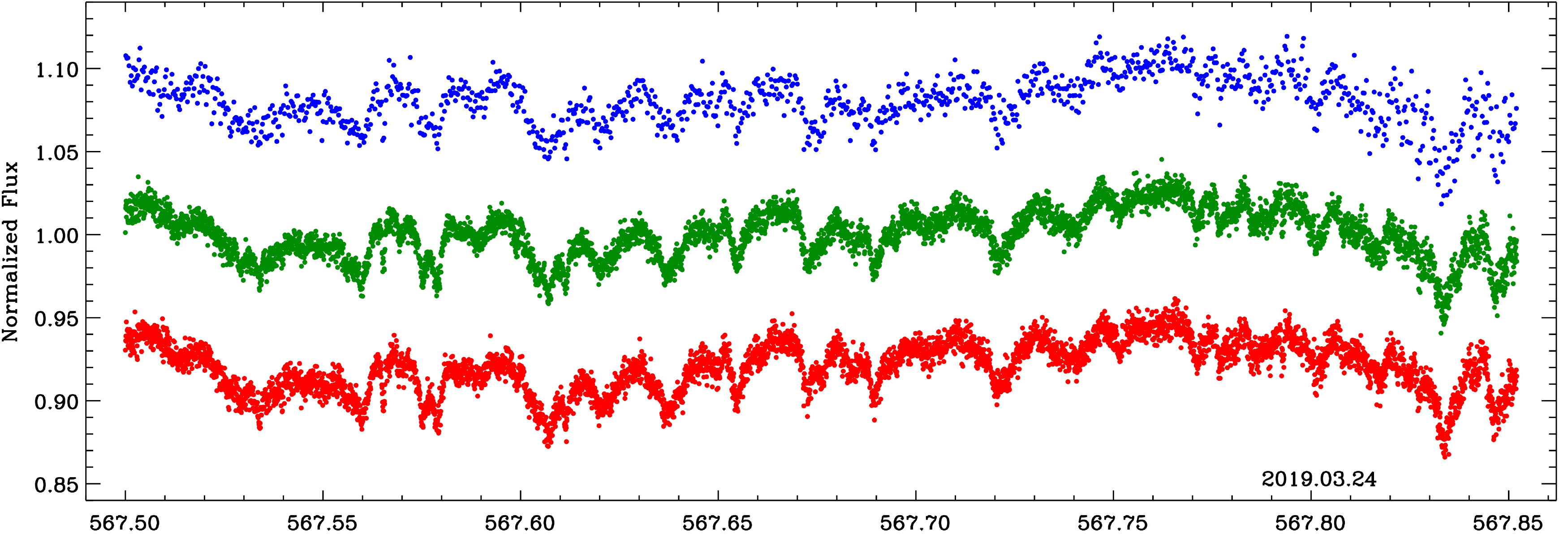}
\includegraphics[width=1.0\linewidth]{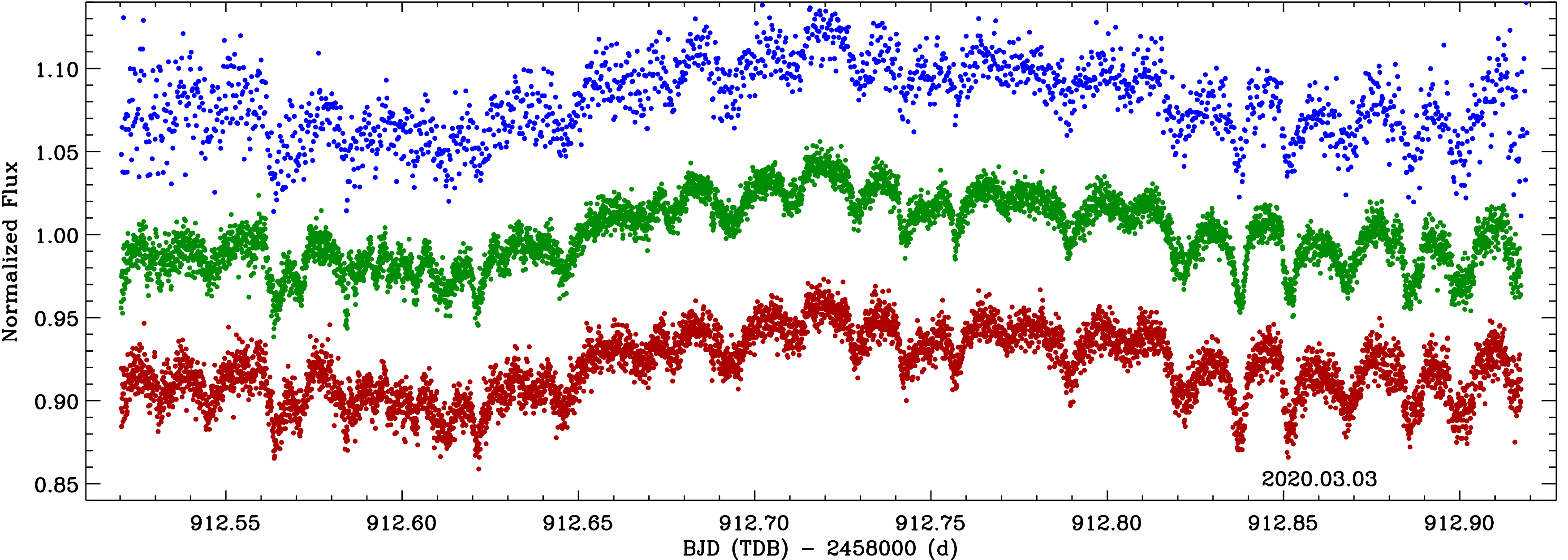}
\caption{ULTRACAM multi-wavelength light curves for WD\,1054--226, taken on the first and latest nights of observation, as labelled in the figure panels.  The data are normalized and offset for clarity, where the blue points are $u$ band, the green points are $g$ band, the red points are $r$ band, and the maroon points are $i$ band.
\label{fig:firstlastcol}}
\end{figure*}

%% FIGURE ALL 18 UCAM NIGHTS %%%
\begin{figure*}
\includegraphics[width=1.0\linewidth]{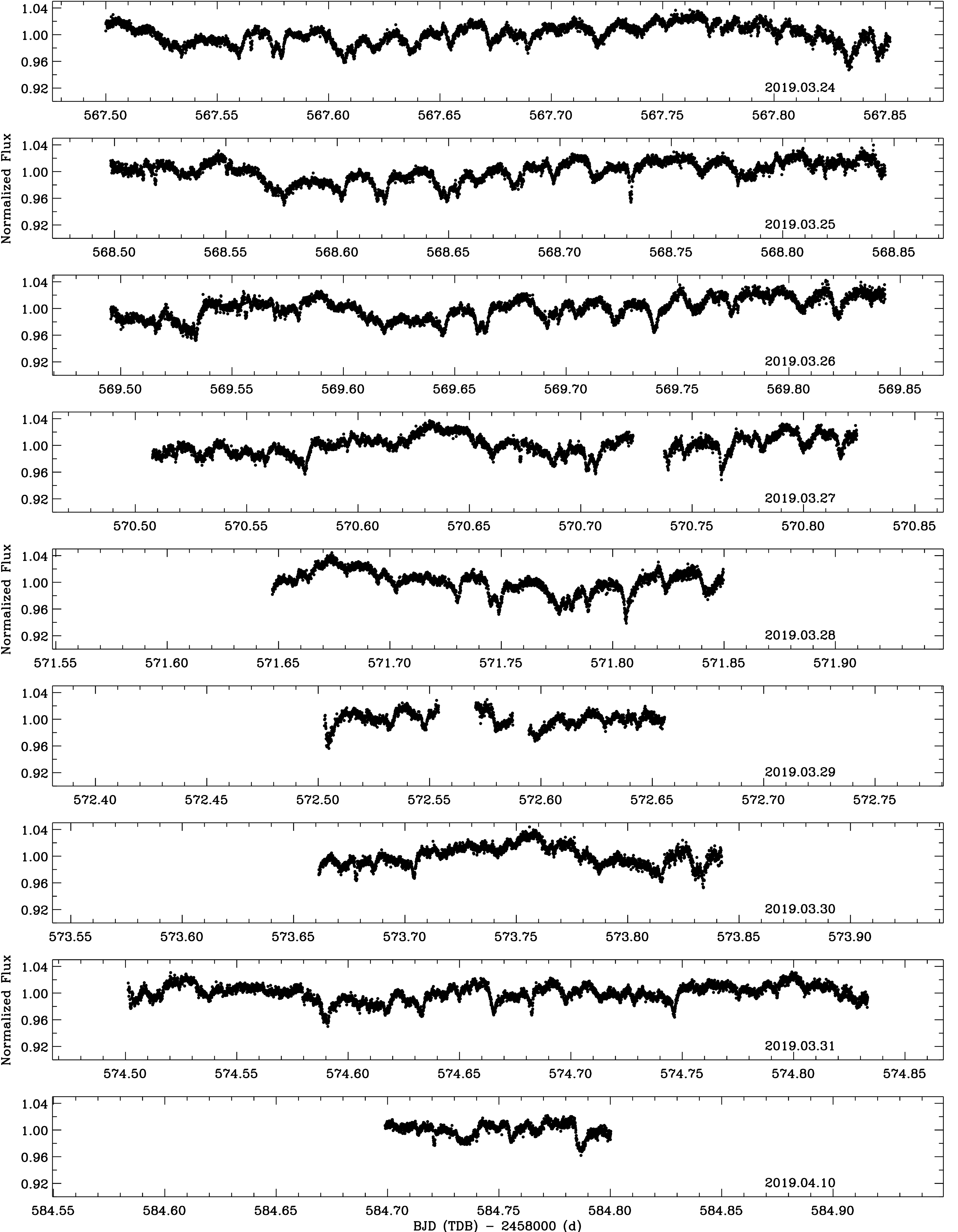}
\caption{All ULTRACAM light curves obtained as detailed in Table~\ref{tbl:obsruns}.  The plotted data are the average of either $g$- and $r$-band, or $g$- and $i$-band observations.
\label{fig:nights18}}
\end{figure*}

%% FIGURE CONT'D %%%
\begin{figure*}
\includegraphics[width=1.0\linewidth]{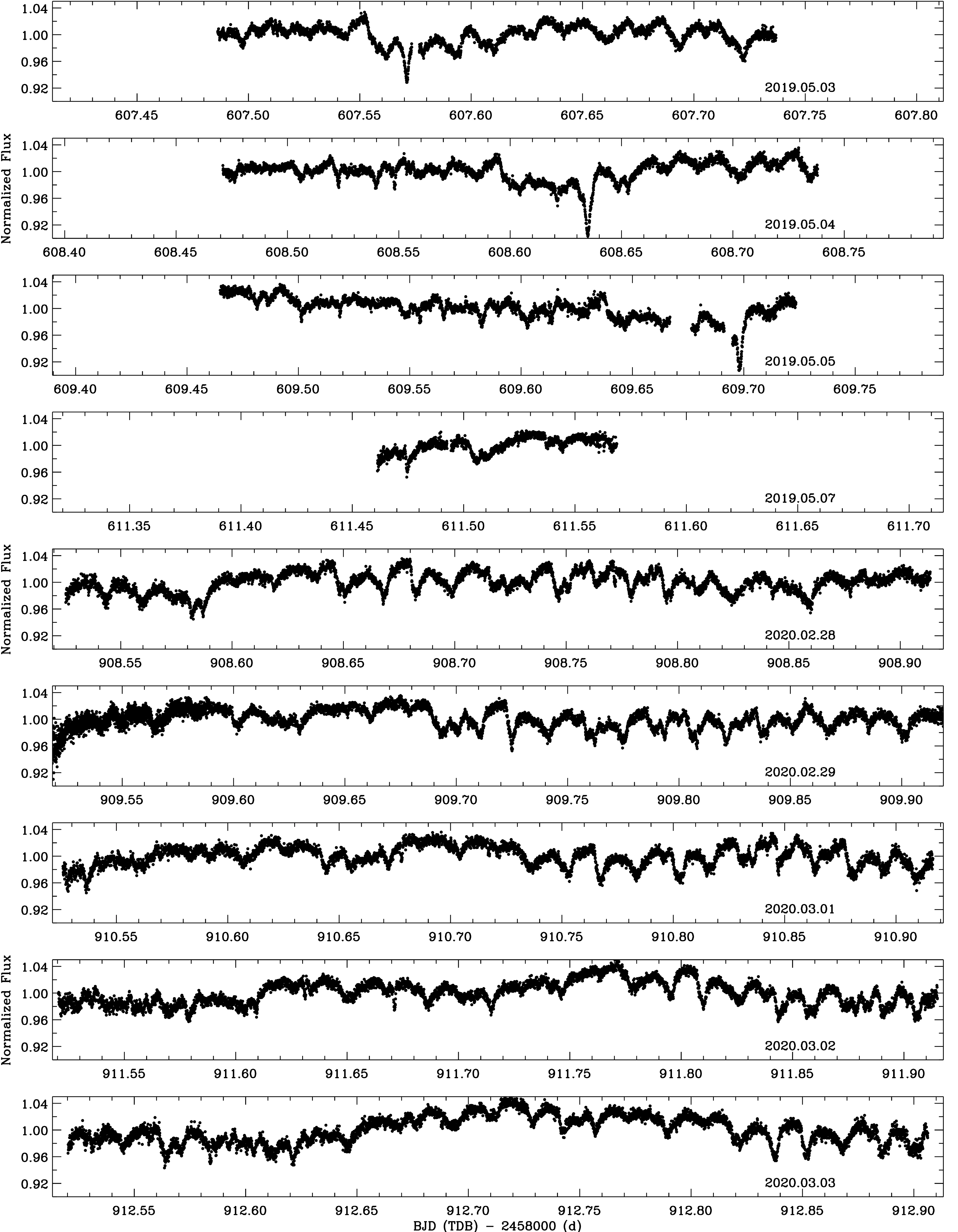}
\contcaption{}
\end{figure*}

%% FIGURE UCAM RESIDUALS %%%
\begin{figure*}
\includegraphics[width=1.0\linewidth]{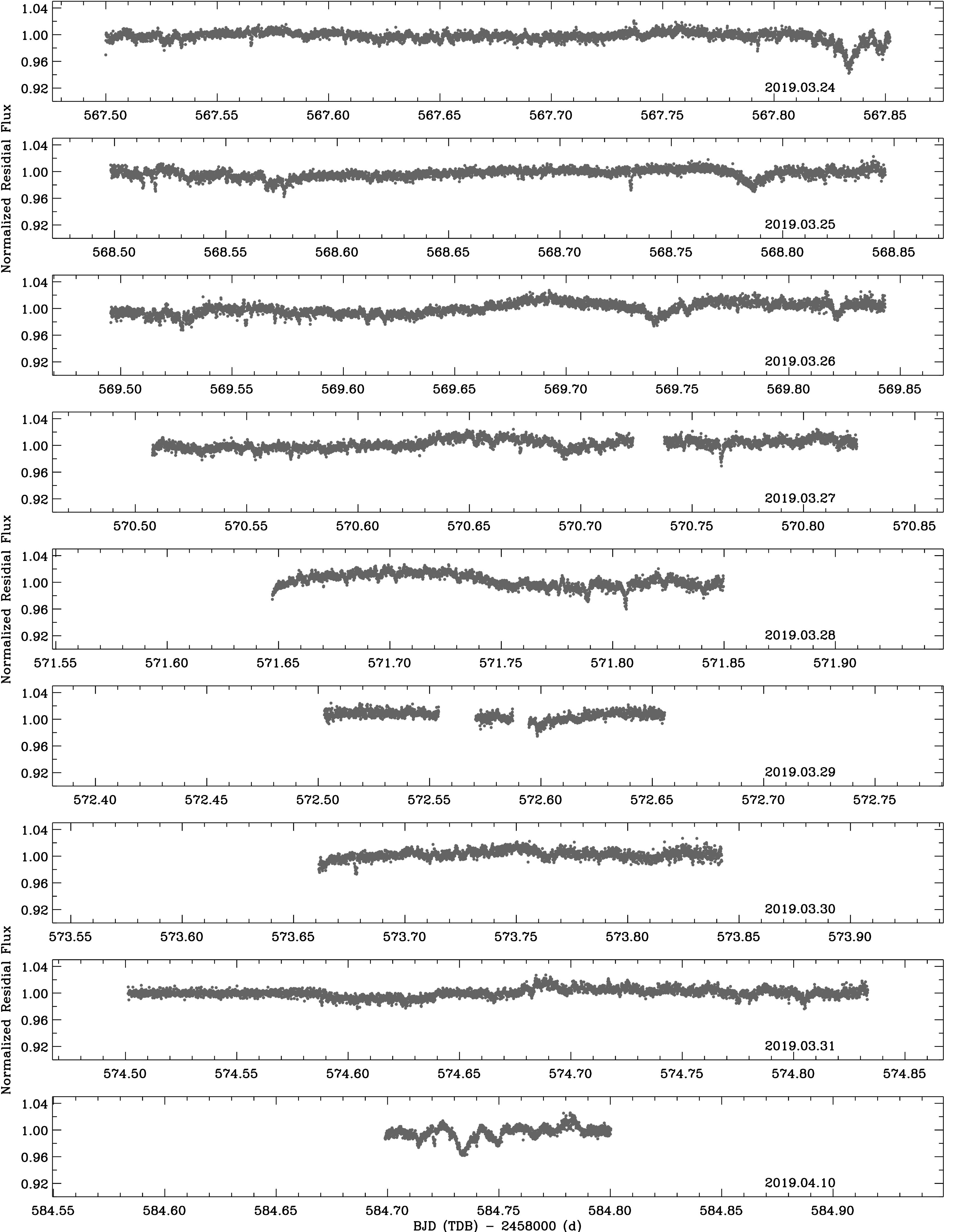}
\caption{All ULTRACAM light curve residuals after removal of phase-averaged data where possible, all performed with an average of either $g$- and $r$-band, or $g$- and $i$-band data.  It is important to note that not all light curve phases are effectively co-added and removed in these residuals, owing to a lack of sufficient coverage at some phases.  This incomplete coverage can introduce artefacts in these light curve residuals, some of which mimic long transits (e.g.\ on 2019 March 31).
\label{fig:resids18}}
\end{figure*}

%% FIGURE CONT'D %%%
\begin{figure*}
\includegraphics[width=1.0\linewidth]{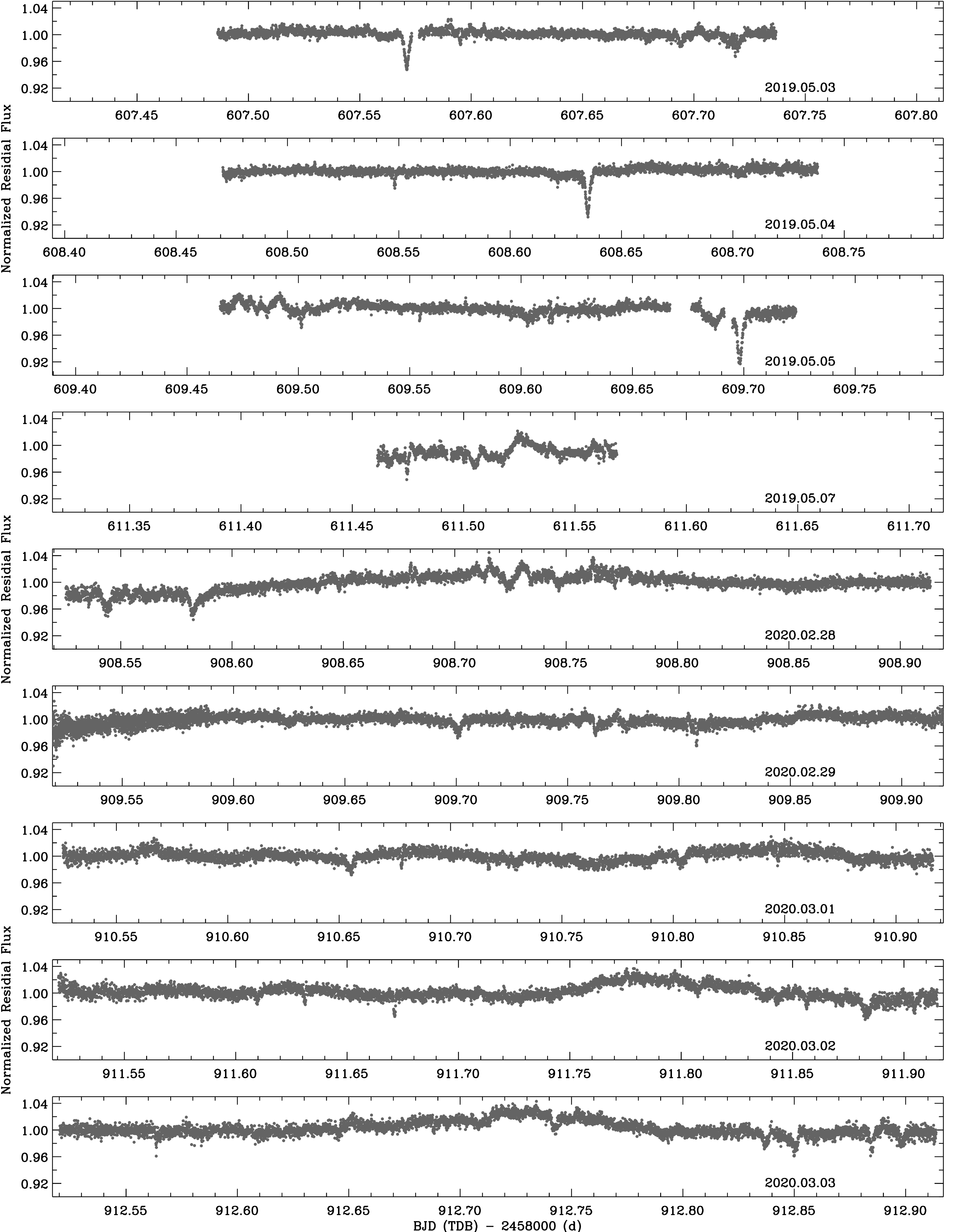}
\contcaption{}
\end{figure*}

\bsp    % typesetting comment
\label{lastpage}
\end{document}